\documentclass{article}

\usepackage{arxiv}

\usepackage[utf8]{inputenc} % allow utf-8 input
\usepackage[T1]{fontenc}    % use 8-bit T1 fonts
\usepackage{hyperref}       % hyperlinks
\usepackage{url}            % simple URL typesetting
\usepackage{booktabs}       % professional-quality tables
\usepackage{amsthm,amsmath,amsfonts}
\usepackage[square]{natbib}
\usepackage{float}
\usepackage{subcaption}
\usepackage{xcolor}
\usepackage{multirow}
\usepackage{nicefrac}       % compact symbols for 1/2, etc.
\usepackage{microtype}      % microtypography
\usepackage{lipsum}
\usepackage{graphicx}
\graphicspath{ {./images/} }

\setcitestyle{numbers}

\title{Nowcasting COVID-19 incidence indicators during the Italian first outbreak}

\author{
    Pierfrancesco Alaimo Di Loro\\
    \scriptsize{Dpt. of Statistical Sciences}\\
    \scriptsize{University of Rome "La Sapienza"}\\
    \scriptsize{\texttt{pierfrancesco.alaimodiloro@uniroma1.it}}
\And
    Fabio Divino\\
    \scriptsize{Dpt. of Bio-Sciences}\\
    \scriptsize{University of Molise}\\
    \scriptsize{\texttt{fabio.divino@unimol.it}}
\And
    Alessio Farcomeni\\
    \scriptsize{Dpt. of Economics and Finance}\\
    \scriptsize{University of Rome "Tor Vergata"}\\
    \scriptsize{\texttt{alessio.farcomeni@uniroma2.it}}
\And
    Giovanna Jona Lasinio \\
    \scriptsize{Dpt. of Statistical Sciences}\\
    \scriptsize{University of Rome "La Sapienza"}\\
    \scriptsize{\texttt{giovanna.jonalasinio@uniroma1.it}}
\And
    Gianfranco Lovison\\
    \scriptsize{Dpt. of Economics, Management and Statistical Sciences}\\
    \scriptsize{University of Palermo}\\
    \scriptsize{Dpt. of Epidemiology and Public Health}\\
    \scriptsize{Swiss TPH Basel}\\
    \scriptsize{\texttt{gianfranco.lovison@unipa.it}}
\And
    Antonello Maruotti\\
    \scriptsize{Dpt. GEPLI}\\
    \scriptsize{Libera Università Maria Ss Assunta}\\
    \scriptsize{Dpt. of Mathematics}\\
    \scriptsize{University of Bergen}\\
    \scriptsize{\texttt{a.maruotti@lumsa.it}}
\And
    Marco Mingione\\
    \scriptsize{Dpt. of Statistical Sciences}\\
    \scriptsize{University of Rome "La Sapienza"}\\
    \scriptsize{\texttt{marco.mingione@uniroma1.it}}
}

\begin{document}
\maketitle
\begin{abstract}
{A novel parametric regression model is proposed to fit incidence data typically collected during epidemics.
The proposal is motivated by real time monitoring and short-term forecasting of the main epidemiological indicators within the first outbreak of COVID-19 in Italy.
Accurate short-term predictions, including the potential effect of exogenous or external variables are provided; this ensures to accurately predict important characteristics of the epidemic (e.g., peak time and height), allowing for a better allocation of health resources over time.
Parameters estimation is carried out in a maximum likelihood framework. All computational details required to reproduce the approach and replicate the results are provided.}
{COVID-19, Growth curves, Richards' equation, SARS-CoV-2, GLM}
\end{abstract}

\section{Introduction}
\label{sec1}

Italy has been the first European country to be severely hit by the first epidemic wave due to the spread of the SARS-CoV-2 virus.
COVID-19 syndrome emerged in northern Italy in February 2020, with a basic reproduction number $R_0$ between 2.5 and 4 \citep{flaxal:2020}. In its most severe form, COVID-19 has two challenging characteristics \cite{Peeri2020}: it is highly infectious and, despite having a benign course in the vast majority of patients, it requires hospital admission and even intensive care for about $10\%$ of infected.
During the outbreak, it was crucial to set up appropriate data collection and modeling systems quickly.
Both were necessary for monitoring infections evolution, evaluation of policy interventions, and prediction.

Generally speaking, the nature of epidemics' spread has nearly always followed the same scenario: first, the growth in the number of infected people is (close to) exponential; in a second moment, this growth gradually but consistently slows down.
So far, in order to explain the spread of epidemics and predict their consequences, a number of mathematical and statistical models of different complexity levels are used. The starting point is often the Verhulst logistic equation \citep{Liang2020}, which can easily capture both the exponential increase in the number of infected people at the initial stage of the epidemic development and the tendency towards a constant value by its ending. In more complex models, people are divided into different groups: (S) the susceptible class, namely those individuals who are capable of contracting the disease and becoming infected; (I) the infected class, namely those individuals who are capable of transmitting the disease to others; (R) the removed class, namely infected individuals who are deceased or have recovered, who are either permanently immune or isolated. This group of mathematical models are called SIR (or compartmental) models (e.g., \cite{diak:13}). References include \cite{chen:20,Giordano2020,Gatto2020,Dehningeabb9789}, and several more.
However, whilst being potentially very appropriate to model the true dynamics underlying any epidemic, the SIR-based models rely on accurate initial estimates of several quantities governing its spreading mechanism (which are mostly unknown). As discussed in \cite{Ioannidis2020}, poor data input on key features of the pandemic can heavily bias these estimates, jeopardizing the reliability of any theory-based forecasting effort.
Indeed, such specifics lead the choice of coefficients in the equations defining the SIR model and define its initial conditions. It is well known that even a slight change in those can lead to large differences in the final results. For instance, at the beginning of the epidemic, early data providing estimates for case fatality rate, infection fatality rate, basic reproductive number, and other key numbers that are essential for the modeling, are often inflated and may cause potentially large over-estimation of the epidemic severity. Similar critiques to using compartmental modeling for nowcasting can also be found in \cite{baek:20}, and references therein.

We have thus preferred to follow an alternative approach, which involved direct modeling of the observed counts (e.g., \cite{Saljeeabc3517}).
We propose a parametric regression model for the modeling of \textit{incidence indicators} (defined in Sec. \ref{Sec_Incidence_Prevalence}) based on the use of the Richard's curve (a generalized logistic function) as response function in place of the widely used exponential or polynomial trend. Furthermore, we replace the generally entrenched Gaussian assumption for the distribution of log-counts \citep{grasselli2020b,sebastiani2020} by the more appropriate Poisson or Negative Binomial distributions for counts. In this way, we avoid the implausible assumptions stemming from the more common alternatives: the former allows the underlying counts to potentially grow indefinitely; the latter neglects the proper specification of dependence between mean and variance under the log-normal distribution.
We further propose different ways of including exogenous information as a linear effect on the response function of counts in a very general generalized linear model framework.
These models have been implemented during the outbreak with the aim of modeling the medium to long term evolution of the epidemic wave.

The use of logistic-based curves is also widely discussed in the literature \citep{cabr:20,greco:20,ritz:15}.
Logistic growth curves can be seen as a flexible formulation for approximating a large variety of growth phenomena, especially in biology and in epidemiology \citep{hsu1984mathematical,grossman1985logistic,morris1992use,berkson1944application,wachenheim2003analysis}. In particular, highly flexible parametric models such as Gompertz curves and the unified-Richards family \citep{TJORVE2010} have been proposed in the study of organisms' growth, for a review see \cite{Tjorve2017}.

The paper is organized as follows:
Section \ref{Sec:ADataLim} gives a detailed description of the Italian situation and provides a brief account of the Italian public data made available daily, with some remarks on limitations and flaws in the data collection process; Section \ref{Sec:ModSpec} contains a description of our approach to modeling incidence indicators, including remarks on how to obtain standard errors for parameters and predictions; Section \ref{Sec:NCItOutbreak} illustrates results of our approach applied to the incidence indicators recorded during first wave of the Italian outbreak of COVID-19. In particular, Sections \ref{Subsec:sahead} and \ref{Subsec:pday} include the evaluation of the (out-of-sample) now-casting accuracy both in terms of future counts predictions and the anticipated forecast of the day of the peak.
Finally, results are discussed and commented along with some concluding remarks in Section \ref{Sec:Duscussion}.

The methods discussed in this paper have also been implemented in a Shiny app, publicly available at \url{https://statgroup19.shinyapps.io/StatGroup19-Eng/}.

\section{Available data and their limitations}
\label{Sec:ADataLim}

The Italian Dipartimento della Protezione Civile (DPC, civil protection department), starting from February 24th, 2020, has been gathering data at the regional level every day and making these public in a {\tt gitHub} repository. During most of the Italian epidemic, data were commented by the department head in an official press release at about 6 p.m. . The daily updated data are currently stored at {https://github.com/pcm-dpc/COVID-19}.

For public health service purposes, Italy is divided into 21 regions. There are 19 administrative regions, plus two autonomous provinces (Trento and Bolzano) that form the administrative region of Trentino-Alto-Adige. We chose to merge (by summing) data about Trento and Bolzano and use the 20 administrative regions as a territorial reference in our analyses.

\subsection{Incidence and prevalence indicators: different mathematical features} \label{Sec_Incidence_Prevalence}

The epidemiological data provided by Protezione Civile can be distinguished into two
basic types:
\begin{enumerate}
  \item \textbf{incidence indicators (flows)}
  \item \textbf{prevalence indicators (stocks)}
\end{enumerate}

\subsubsection{Incidence indicators} \label{Subsubsec:Incidence}

\begin{figure}[t]
    \centering
    \includegraphics[width=0.8\textwidth]{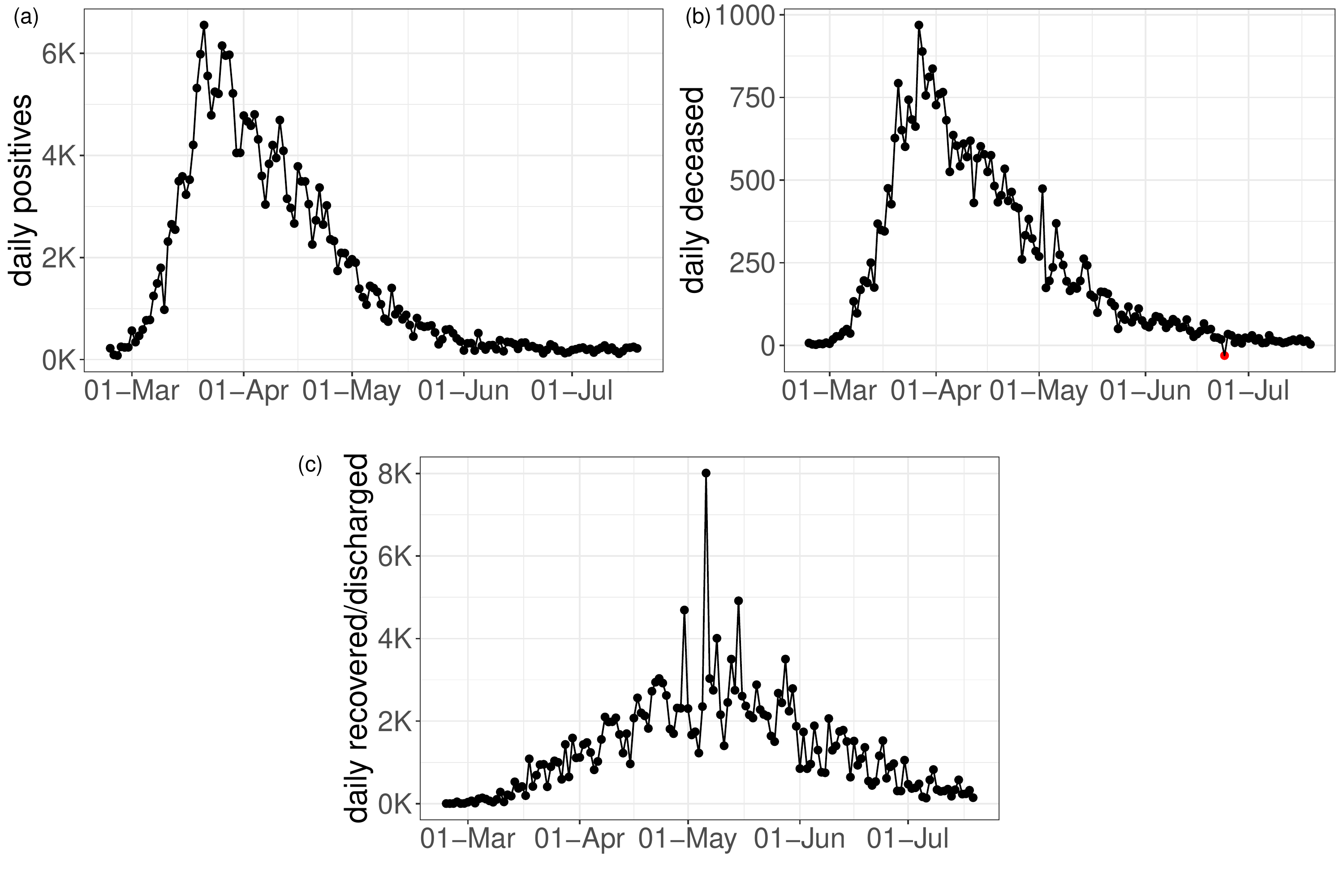}
     	\caption{Time series of the Italian daily incident indicators: \textit{daily positives} (a), \textit{daily deceased} (b) and \textit{daily recovered/discharged} (c).}
	\label{DailyIncidence}
\end{figure}

Incidence indicators measure the number of individuals with a particular condition, related with the epidemic, recorded during a given period.
They can be referred to different time periods;
in particular, in the Protezione Civile dataset, \textbf{daily incidence counts} are available for the
following indicators:
\begin{itemize}
  \item positives, which are sub-classified into two sub-conditions:
  \begin{itemize}
    \item hospitalised (either in regular wards or in ICU)
    \item isolated-at-home
  \end{itemize}
  \item deceased
  \item recovered/discharged
\end{itemize}
These indicators can be considered, by analogy with the terminology used in econometrics, as \textbf{flow data}, quantifying the daily input (positives) and output (deceased
and recovered/discharged) of the system. Fig. \ref{DailyIncidence} shows the time series of daily incidence indicators.

\begin{figure}[t]
    \centering
    \includegraphics[width=0.8\textwidth]{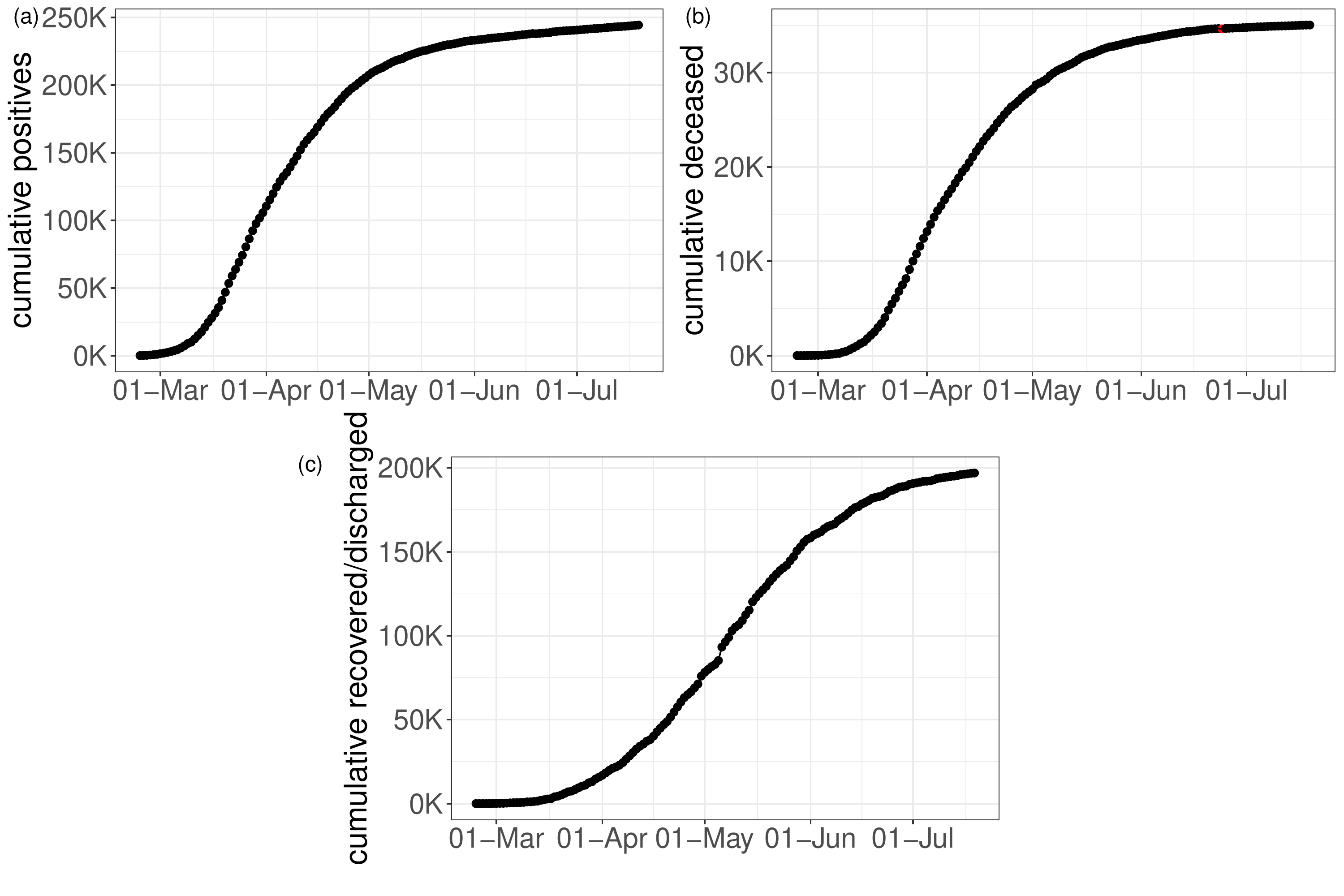}
     \caption{Time series of the Italian cumulative incident indicators: \textit{cumulative positives} (a), \textit{cumulative deceased} (b) and \textit{cumulative recovered/discharged} (c).}
   	\label{CumPosDec}
\end{figure}

One important feature of these indicators, from the viewpoint of the following modeling effort, is that they can be referred to longer time intervals, simply aggregating them over time. The most interesting \textbf{cumulative incidence indicators}
are those referring to the whole history of the pandemic, computed from a conventional
date of "beginning of the pandemic" (typically, the day the systematic recording of daily
positives began) to the current day:
\begin{itemize}
  \item cumulative positives
  \item cumulative deceased
  \item cumulative recovered/discharged
\end{itemize}
By their nature of cumulative counts, these data series are necessarily monotonically non-decreasing. Fig. \ref{CumPosDec} shows the time series of cumulative incidence indicators.

\subsubsection{Prevalence indicators} \label{Subsubsec:Prevalence}

Prevalence indicators measure the number of individuals with a particular condition, related with the epidemic, at a given instant in time (or at a given short interval of time, e.g. a day).
They are typically obtained from simple algebra from other indicators; in particular, in the Protezione Civile dataset, the
following indicators are available daily:
\begin{itemize}
  \item current positives ($CP$)
  \item current ICU occupancy ($ICU$)
\end{itemize}
Again, these indicators can be considered, by analogy with the terminology used in econometrics,
as \textbf{stock data}, resulting from the balance between total inputs (cumulative positives, etc.) and outputs
(cumulative deceased and recovered/discharged, etc.) of the system.

\begin{figure}[t]
    \centering
    \includegraphics[scale=0.4]{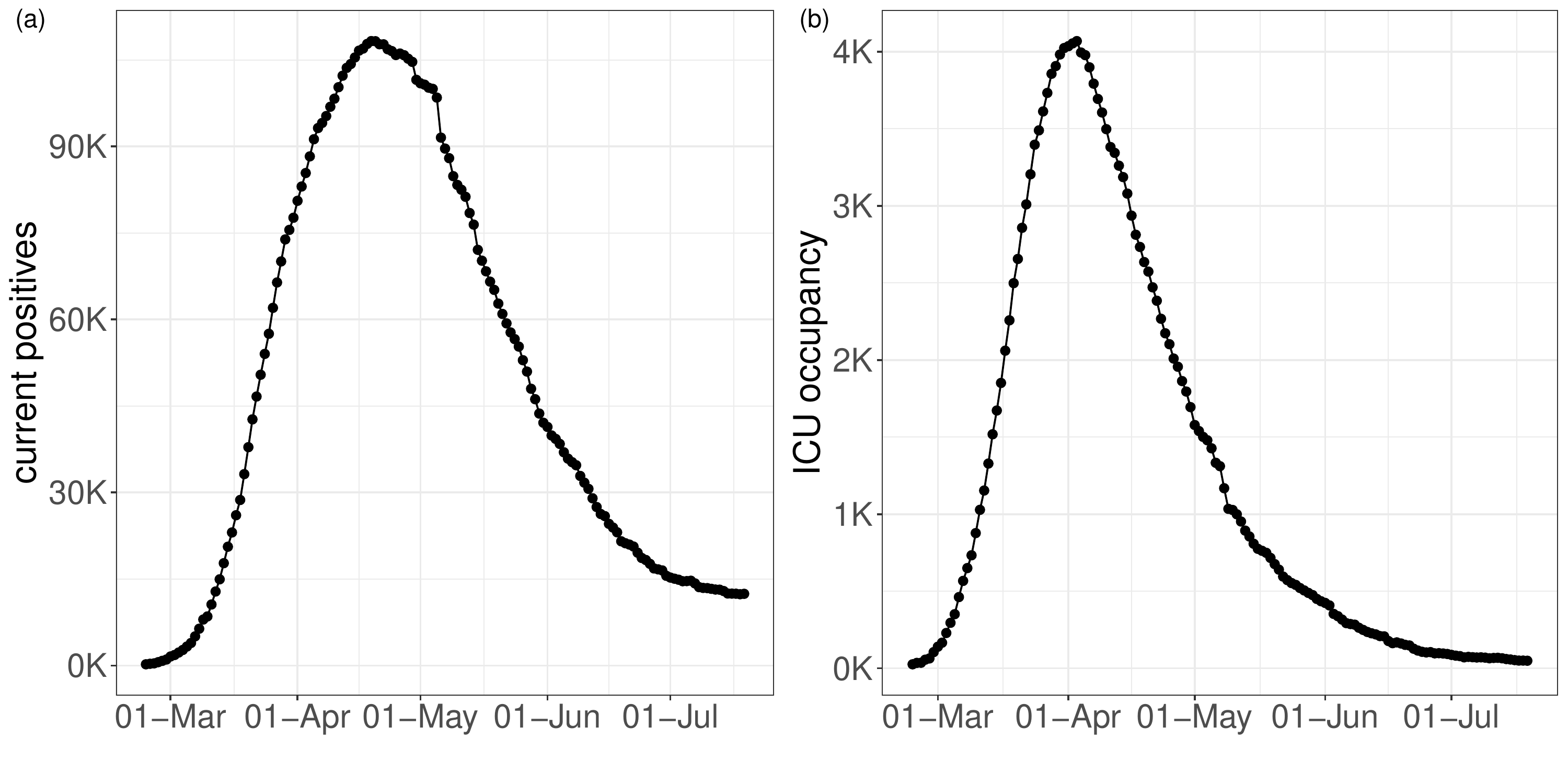}
     \caption{Time series of Italian daily prevalence indicators: \textit{current positive} (a) and \textit{ICU occupancy} (b).}
   	\label{DailyPrevalence}
\end{figure}

Two important features of these indicators are that:
\begin{enumerate}
  \item given their \textit{stock} nature, they cannot be aggregated (e.g.: it does not make sense to compute "cumulative current positives");
  \item by their own nature, these indicators are not monotone, since they can increase or decrease as a result of different trends of the component series. Typically, we expect the series of current positives and ICU occupancy to increase in the rising phase of an epidemic, reach a peak
  and then decrease to a lower asymptote (see Fig. \ref{DailyPrevalence}), although more complex patterns due to resurgence of the epidemic are also plausible.
\end{enumerate}

Prevalence indicators are characterized by a strong and tangled dependence structure which is cumbersome to simplify into a manageable and useful statistical model on the short run. \\
For this reason, the focus of this work concerns only incidence indicators. Our model proposal, from a strictly mathematical point of view, could potentially applied also on prevalence indicators. However, from the statistical point of view, the modeling assumptions which are assumed to hold (with good approximation) considering the incidence indicators, are likely to be strongly violated by prevalence indicators and the resulting outcome cannot be considered reliable. In Sec. \ref{Sec:Duscussion} we shall briefly discuss some possible
approaches for the analysis of prevalence indicators. 

\subsection{Data issues}
\label{sec:DataIss}

COVID-19 public Italian data present several issues that severely affect their quality.
The information has been gathered and reported at a regional level, and each regional healthcare organization has a different transmission and data collection system\footnote{see \url{https://www.epiprev.it/materiali/2020/EP2-3/112_edit1.pdf} for further details.}. 

Measurement errors, and errors in data entry, are expected to be often present. Delays in reporting has been, sometimes, substantial.
Some patients were transferred (e.g., from Lombardia to Puglia, and even to Germany) without notification, and they were counted as hospital patients of the receiving region (or not at all when sent abroad) and positive cases of the region of residence.
Most importantly, counts were updated on the notification day rather than aligned to a more appropriate date. For example, death is counted on the day of the reporting, not on the day of the outcome, which could be even weeks before. Positive status is also counted on the day that test results are received, with swabs being done from one day to weeks after symptoms' onset. No distinction between actively symptomatic and asymptomatic patients was made.\\
Swabs and positive cases are not time-aligned. For example, in countries like Singapore  (\url{https://www.moh.gov.sg/covid-19}), daily data include information on total swabs tested, total unique persons swabbed as well as total swabs per 1,000,000 total population and total unique persons swabbed per 1,000,000 total population.
In Italy, up to the $19$-th of April $2020$, only the total number of daily swabs is available, and no linkage between swabs and tested individuals was kept in the data repository. Hence, it is impossible to make statistically sound use of swabs' count to model the whole first pandemic wave.\\
%On the other hand, the disentangled information on tested individuals is available, daily, from the same date on. An approach at including such information is presented in Appendix \ref{sec:modtested}.\\
Finally, it is crucial to recall that people diagnosed with COVID-19 disease are only a small fraction of the people infected by the virus. Moreover, since the tracking was highly symptoms' driven, especially in the first phase of the outbreak, the detected number of positives cases can provide only a partial estimate of the \textit{true} incidence of COVID-19 in the Italian population. Eventually, we expect this detected fraction to vary wildly over space and time. \\
In our opinion, the most reliable indicator is the count of ICU occupancy. The reason is that the Italian Society for Emergency Care issued national guidelines (that did not change substantially during the epidemic) for testing patients with a suspected infection by SARS-CoV-2, who also had top priority for swab access and reporting; and ICU admissions can be expected to depend on the proportion of infected population susceptible to severe infection, rather than to the regional strategy for testing and contact tracing.
However, while probably reliable, this indicator does present some drawbacks.
First of all, it provides only a partial snapshot of the epidemic's current stage, which concerns the most severe cases of the disease.
The latter is a critical issue, especially in the COVID-19 case, which is known to present severe symptoms only in a small percentage of the currently affected individuals.
Second, this snapshot is affected by a constant delay (i.e., the time between catching the disease and manifesting severe symptoms).
As mentioned in Sec. \ref{Subsubsec:Prevalence}, its daily variation is obtained as a combination of new incoming patients (+) and the deceased or recovered ones (-), whose effects blend and are hard to disentangle.
As a consequence, \textit{incidence indicators}, such as \textit{daily positives} and \textit{daily deceased}, while being measured with some error and even more delay in the case of deaths, still represent the critical indicators for timely and appropriate monitoring of the pandemic.

\section{Model specification}
\label{Sec:ModSpec}

The time series of any of the observed indicators, denoted by $\mathbf{z}=\left\lbrace z_t\right\rbrace_{t=t_0}^T$, is
modeled separately and considered as the realization of the stochastic process $\mathbf{Z}=\left\lbrace Z_t\right\rbrace_{t=t_0}^T$.
The idea behind this paper is to model any of the mentioned indicators through a Generalized Linear Model with a response function $\mathbb{E}[Z_t] = \mu(t) = g^{-1}\left(t; \boldsymbol{\gamma}\right)$, where $g(\cdot)$ is a known link function and $\boldsymbol{\gamma}$ is a parameter vector, that is appropriate for the specific mathematical features of the epidemic process. This must be coupled with a response distribution $f(Z_t; \mathbf{\theta})$ coherent with the domain of such indicators, which are counts and therefore Natural numbers.

\subsection{Response function for incidence indicators} \label{Subsec:RespInc}

\begin{figure}[t]
    \centering
    \includegraphics[scale=0.4]{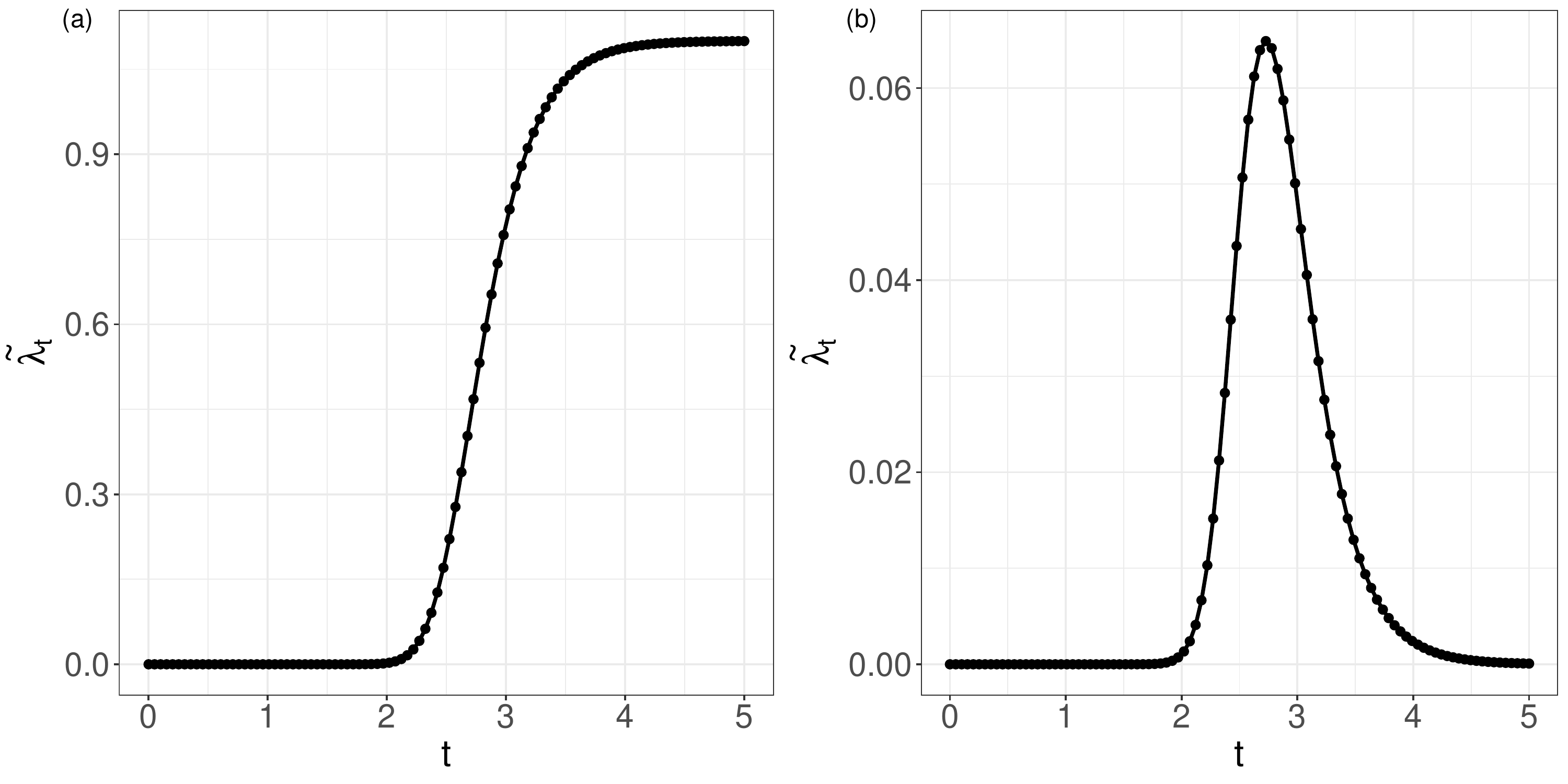}
     \caption{Example of Richard's curve (a) and derivative of the Richard's curve (b).}
   	\label{RichardsExamples}
\end{figure}

Let us denote by $\left\lbrace y^c_t\right\rbrace_{t=0}^T$ the time-series of cumulative incidence indicators since the start of the epidemic ($t_0 = 0$, first day of systematic data recording).
Visual inspection of these indicators in Fig. \ref{CumPosDec} suggests that their expected values follow a logistic-type growth curve.
For these indicators, we consider a \textit{Generalized Logistic Function}, also known  as \textit{Richards' Curve}
(see Fig. \ref{RichardsExamples} as an example), as response function for the mean of the process \citep{richards1959flexible}.
Richards' response function depends on $5$ parameters $\boldsymbol{\gamma}^T = [b, r, h, p, s]$ and can be expressed as:
\begin{equation}
\label{expRich}
    \mathbb{E}[Y^c_t] = g^{-1}(t; \boldsymbol{\gamma}) = \lambda_{\boldsymbol{\gamma}}(t) = b + \frac{r}{(1 + 10^{h(p-t)})^s}
\end{equation}
where $b\in\mathbb{R}^+$ represents a lower asymptote and $r>0$ is the distance between the upper and the lower asymptote. The parameter $h$ is the \emph{hill} (growth rate), $p\in (0,T)$ is a peak position parameter: it tells when the curve growth speed slows down, and $s\in\mathbb{R}$ is an asymmetry parameter. In our context, since cumulative incidences are always monotone increasing indicators, it is reasonable to assume $h,s>0$\footnote{Conversely, we may assume $h,s<0$}.
An extensive review of Gompertz models and proper interpretation of the parameters is given in \cite{TJORVE2010}.

An extended GLM with (\ref{expRich}) as response function seems to be a natural choice for modeling time series of \textit{cumulative counts}, whose monotonically non-decreasing average behaves as the \textit{Richard's curve}. Unfortunately, there is a significant drawback to this choice. As it will be better clarified in the Sec \ref{RespDistriIncInd}, a very useful working assumption would be that all these counts were stochastically independent, given their mean function $\lambda_{\boldsymbol{\gamma}}(t)$.
However, we cannot consider this assumption as realistic in the case of cumulative counts, since the constraint on the domain of definition on subsequent counts (i.e., $y^c_t \geq y^c_{\tau}, \forall \tau < t$) is not guaranteed to be satisfied.
On the other hand, the stochastic independence assumption sounds more reasonable, albeit not necessarily true, for the \textit{daily incidence counts} $\left\lbrace y_t\right\rbrace_{t=1}^T$, i.e., the addenda of the cumulative counts excluding the starting point $y_{0}$, which can be defined as:
\begin{equation*}
    y^c_t= \sum_{\tau=0}^t y_\tau \quad\Rightarrow\quad y_t= y^c_t-y^c_{t-1},\qquad t=1,\dots,T
\end{equation*}
where $y_0=0$ by definition.\\
Using Equation \ref{expRich}, and exploiting the additive properties of the expected value, we have:
\begin{equation*}
\label{eqRespFun}
    \begin{aligned}
    \tilde{\mu}(t) = \mathbb{E}[Y_t] &= \mathbb{E}[Y^c_t] - \mathbb{E}[Y^c_{t-1}] =  \lambda_{\boldsymbol{\gamma}}(t) - \lambda_{\boldsymbol{\gamma}}(t-1) = \\
    &= r \cdot \left[(1 + 10^{h(p-t)})^{-s}-(1 + 10^{h[p-(t-1)]})^{-s}\right]=\tilde{\lambda}_{\boldsymbol{\gamma}}(t)
    \end{aligned}
\end{equation*}
which, in particular, does not depend on the baseline $b$. Therefore, we shall adopt an extended GLM with response function given by the first differences of the Richards Curve $\boldsymbol{\tilde{\lambda}_{\boldsymbol{\gamma}}}=\left\lbrace\tilde{\lambda}_{\boldsymbol{\gamma}}(t)\right\rbrace_{t=1}^T$ to model the daily expected values $\boldsymbol{\mu}=\left\lbrace\mu(t)\right\rbrace_{t=1}^T$ of the observed incidence counts $\mathbf{y}=\left\lbrace y_t \right\rbrace_{t=1}^T$ (see example in Fig. \ref{RichardsExamples}).

In addition, we may also consider adding a kink effect/baseline $\alpha$ to the first differences $\tilde{\lambda}_{\boldsymbol{\gamma}}(\cdot)$, which is to say assuming the following functional form for the mean of the daily counts:
\begin{equation}
\label{eqRespFunBase}
        \tilde{\mu}_{\boldsymbol{\theta}}(t) = \alpha + \tilde{\lambda}_{\boldsymbol{\gamma}}(t), \quad \alpha\geq 0,
\end{equation}
where $\boldsymbol{\theta}=(\alpha,\boldsymbol{\gamma})$. This would correspond to the following mean function for the cumulative counts:
\begin{equation*}
    \mu_{\boldsymbol{\theta}}(t) = \alpha\cdot (t-1) + \lambda_{\boldsymbol{\gamma}}(t).
\end{equation*}
In practice, the parameter $\alpha$ includes the possibility of having a strictly positive baseline rate, which can be interpreted as the \textit{endemic steady state incidence rate}. On the other hand, the first differences of the Richard's Curve $\tilde{\lambda}_{\boldsymbol{\gamma}}(t)$ are (by construction) forced to decrease asymptotically to the value of $0$.
However, this asymptotic result is not necessarily observed in real data. In particular, Fig. \ref{DailyIncidence} highlights that both time-series do not attain the $0$ value, but settle to a low, constant level.
This situation may, potentially, continue indefinitely: new cases will be found as long as people will be tested.
Consequently, the model without a baseline lacks the ability to catch this tail and, because of the curve parametric form, this may indirectly affect the fit on the whole series.

In the first instance, one solution would be to fit the model, including the kink effect $\alpha$. Afterward, if it is estimated to be sensibly different from $0$, the model without $ \alpha $ can be fitted again to stabilize the estimation procedure and decrease the uncertainty on the other parameters.

\subsection{Response distribution for incidence indicators}\label{RespDistriIncInd}

Before introducing the distributions for the daily incidence counts, we must make some assumptions about the time dependence structure. In particular, we assume that given the mean function $\tilde{\lambda}_{\boldsymbol{\gamma}}(\cdot)$, the daily incidence counts $Y_t$ are:
\begin{itemize}
    \item stochastically independent from the previous state $Y^c_{t-1}$ given the mean function $\lambda_{\boldsymbol{\gamma}}(\cdot)$:
    \begin{equation*}
        Y_t \perp Y^c_\tau\;\forall\;\tau< t
    \end{equation*}
    \item stochastically independent among them: $$Y_t\perp Y_\tau\; \forall t,\tau$$
\end{itemize}
This implies a $1$-st order Markov property for the cumulative counts:
\begin{equation*}
Y^c_t|Y^c_{t-1}\perp Y^c_{1:t-2}\;\forall t
\end{equation*}
and we can write:
\begin{equation*}
\begin{aligned}
    f_{Y^c_{1:T}}(y^c_1, \dots, y^c_T|y_0; \theta) &= \prod_{t=1}^T f_{Y^c_t}( y^c_t| y^c_{0:t-1}; \theta)
                                                    = \prod_{t=1}^T f_{Y^c_t}(y^c_t| y^c_{t-1}; \theta)=\\
                                                   &=\prod_{t=1}^T f_{Y_t}(y_t| y^c_{t-1}; \theta)=\prod_{t=1}^T f_{Y_t}(y_t| \theta)
\end{aligned}
\end{equation*}
where we recall that the last equivalence is justified by the assumption that, given the parameters, the daily incidence counts are independent w.r.t. to the previous observed cumulative count.

We remark that although the first-order Markov property for the cumulative counts is an approximation in the present case, this kind of approach has been valid for all incidence indicators provided by the Italian Protezione Civile.

For communication purposes, it can be of interest to report the results of analyses and predictions in terms of
cumulative, rather than daily, incidence indicators. Clearly, it is
possible to model and predict the daily incidence indicators and, from these estimates and predictions, obtain the relevant cumulative incidence indicators.

\subsubsection{Poisson distribution}

Let us assume that the vector of daily incident counts, $\mathbf{y}=\left\lbrace y_1,\dots,y_t\right\rbrace$, is composed of independent Poisson realizations with expected value $\tilde{\mu}_{\boldsymbol{\theta}}(t)$:
\begin{equation*}
\label{PoisNonInd}
    Y_t|\boldsymbol{\theta} \sim Pois(\tilde{\mu}_{\boldsymbol{\theta}}(t)), \quad t = 1, \dots, T
\end{equation*}
Hence, the likelihood can be written as:
\begin{equation*}
\begin{aligned}
    \mathcal{L}(\boldsymbol{\theta}|\mathbf{y})&=\prod_{t=1}^T Pois(y_t|\tilde{\mu}_{\boldsymbol{\theta}}(t))\propto\\
    &\propto \tilde{\mu}_{\boldsymbol{\theta}}(t)^{\sum_{t=1}^T y_t} e^{-\sum_{t=1}^T\tilde{\mu}_{\boldsymbol{\theta}}(t)}
\end{aligned}
\end{equation*}
and the log-likelihood is given by:
\begin{equation*}
    \log\mathcal{L}(\boldsymbol{\theta}|\mathbf{y})\propto \sum_{t=1}^T y_t\log\left(\tilde{\mu}_{\boldsymbol{\theta}}(t)\right) - \sum_{t=1}^T\tilde{\mu}_{\boldsymbol{\theta}}(t)
\end{equation*}
Remark that, under the assumption of Poisson distribution and the baseline $\alpha=0$ (i.e. $\tilde{\mu}_{(\alpha, \boldsymbol{\gamma})}=\tilde{\lambda}_{\boldsymbol{\gamma}}(\cdot)$), we can exploit the well-known Poisson's additive property\footnote{the sum of independent Poissons is still a Poisson with parameter the sum of the parameters} to conclude that each cumulative count $Y^c_t$ is still marginally distributed according to a Poisson, parametrized by the original Richard's Curve function $\lambda_{\boldsymbol{\gamma}}(\cdot)$:
$$
Y^c_t|\boldsymbol{\gamma} \sim Pois\left(\sum_{\tau=1}^t\tilde{\lambda}_{\boldsymbol{\gamma}}(\tau)\right)= Pois(\lambda_{\boldsymbol{\gamma}}(t))
$$

\subsubsection{Negative Binomial distribution}

When counts are over-dispersed the Poisson distribution is not a suitable choice. We can model the observed daily incidence counts $\mathbf{y}=\left\lbrace y_1,\dots,y_t\right\rbrace$ as independent realizations from a Negative Binomial with mean $\tilde{\lambda}_{\boldsymbol{\gamma}}(t)$ and dispersion parameter $\nu\in\mathbb{R}^{+}$:
\begin{equation*}
\label{BinNonInd}
    Y_t|\boldsymbol{\theta} \sim NB(\tilde{\mu}_{\boldsymbol{\theta}}(t), \nu), \quad t = 1, \dots, T
\end{equation*}

Hence, the likelihood can be written as:
\begin{equation*}
\begin{aligned}
    \mathcal{L}(\boldsymbol{\theta}, \nu|\mathbf{d})&=\prod_{t=1}^T NB(y_t|\tilde{\mu}_{\boldsymbol{\theta}}(t), \nu)\propto\\
    &\propto \prod_{t=1}^T\left[\frac{\Gamma(\nu+y_t)}{\Gamma(\nu)}\left(\frac{\nu}{\nu+\tilde{\mu}_{\boldsymbol{\theta}}(t)}\right)^\nu\left(\frac{\tilde{\mu}_{\boldsymbol{\theta}}(t)}{\nu+\tilde{\mu}_{\boldsymbol{\theta}}(t)}\right)^{y_t}\right]
\end{aligned}
\end{equation*}
and the log-likelihood is:
\begin{equation*}
\begin{aligned}
	\log\mathcal{L}(\boldsymbol{\theta}, \nu|\mathbf{y})& \propto \sum_{t=1}^T \log\left(\frac{\Gamma(\nu+y_t)}{\Gamma(\nu)}\right) + \nu\sum_{t=1}^T\log\left(\frac{\nu}{\nu+\tilde{\mu}_{\boldsymbol{\theta}}(t)}\right) \\
	&+ \sum_{t=1}^T y_t \log\left(	\frac{\tilde{\mu}_{\boldsymbol{\theta}}(t)}{\tilde{\mu}_{\boldsymbol{\theta}}(t)+\nu}\right)
\end{aligned}
\end{equation*}
The Negative Binomial does not satisfy the same additive property as the Poisson, hence we cannot draw the same conclusion reached in the Poisson case about the marginal distribution of the cumulative count $Y^c_t$ when $\alpha=0$.
In general, the cumulative count in the NB case will follow the distribution stemming from the sum of independent Negative Binomial r.v. with common dispersion parameter $\nu$ but different means $\boldsymbol{\tilde{\mu}}=\left\lbrace\tilde{\lambda}_{\boldsymbol{\gamma}}(t)\right\rbrace_{t=1}^T$.

\subsection{Response function depending on covariates} 
\label{Subsec:IncCovs}

The trend of any of the considered indicators may also depend on additional exogenous information, which we may assume to be known \textit{a priori} either because it is immutable (i.e., the day of the week), or because policy makers fixed it (daily number of \textit{tested cases}/\textit{swabs} set by the government).
For instance: one might want to correct for possible weekly seasonality, which is known to affect the \textit{daily positives} series since many laboratories are closed during the weekend and can not evaluate swabs.
The latter can be used to disentangle the underlying trend of the epidemic from the obvious positive correlation between \textit{tested cases} and \textit{daily positives}.

In general, we may want to include the effect of any set of $k$ time-varying covariates $\underset{T \times (k+1)}{\mathbf{X}} =\left[\mathbf{x}(t)\right]_{t=1}^T$ in the \textit{Richards' GLM} framework through the usual linear predictor:
\begin{equation*}
    \eta(\mathbf{X})=\mathbf{X}\boldsymbol{\beta}
\end{equation*}
where $\boldsymbol{\beta}$ is a $k+1$-dimensional vector of real valued parameters (including intercept).
Let us denote the mean function of the considered indicator as $\tilde{\mu}_\theta(t)=\mathbb{E}\left[Y_t\right]$, where $\boldsymbol{\theta}=(\alpha, \boldsymbol{\gamma}, \boldsymbol{\beta})$.
In order to respect the positivity of the mean parameter (which is necessary both in the Poisson and in the Negative Binomial case), we consider the link function $g(\cdot)=\log(\cdot)$, so that the effect on the mean is expressed as:
\begin{equation*}
    \hat{\eta}(\mathbf{X})=\exp\left\lbrace\eta(\mathbf{X})\right\rbrace=\exp\left\lbrace\mathbf{X}\boldsymbol{\beta}\right\rbrace.
\end{equation*}
Considering a single time point $t$, we would get the following functional form:
\begin{equation*}
    \hat{\eta}(\mathbf{x}(t))=\exp\left\lbrace\mathbf{x}(t)\boldsymbol{\beta}\right\rbrace.
\end{equation*}
Now, the transformed predictor can unload its effect on the mean function either in an additive or a multiplicative fashion.

\subsubsection{Additive inclusion of covariates}
\label{Subsec:addCovs}
The inclusion of an additive effect of covariates implies that the effect of every covariate is constant through-out the pandemic, notwithstanding the current contagion level:
for instance, one may think that an increase of daily \textit{tested cases} will always produce the same increase of daily \textit{daily positives}.
If that is the case, we may just express the baseline parameter $\alpha$ at each time-point $t$ as the \textit{linked} linear combination of covariates $\hat{\eta}(\mathbf{x}(t))$, which would produce the following mean function:
\begin{equation*}
    \tilde{\mu}_\theta(t) = \exp\left\lbrace\mathbf{x}(t)\boldsymbol{\beta}\right\rbrace + \tilde{\lambda}_{\boldsymbol{\gamma}}(t)
\end{equation*}
On the whole vector of observations, this can be expressed as:
\begin{equation*}
    \tilde{\boldsymbol{\mu}}_{\boldsymbol{\theta}} = \exp\left\lbrace\mathbf{X}\boldsymbol{\beta}\right\rbrace + \tilde{\boldsymbol{\lambda}}_{\boldsymbol{\gamma}}.
\end{equation*}

\subsubsection{Multiplicative inclusion of covariates}
\label{Subsec:multCovs}
The inclusion of a multiplicative effect of covariates would imply that the more serious the pandemic situation, the more severe the impact of any covariate on the indicators' daily rate.

First, let us recall that in Sec. \ref{Subsubsec:Incidence} we computed the first differences of the Richard's curve function as:
\begin{equation*}
    \tilde{\lambda}_{\boldsymbol{\gamma}}(t)=r \cdot \left[(1 + 10^{h(p-t)})^{-s}-(1 + 10^{h[p-(t-1)]})^{-s}\right]=r\cdot \tilde{\lambda}_{\boldsymbol{\gamma},-r}(t)
\end{equation*}
where $\tilde{\lambda}_{\boldsymbol{\gamma},-r}(t)=\left[(1 + 10^{h(p-t)})^{-s}-(1 + 10^{h[p-(t-1)]})^{-s}\right]$.
On the log-scale, it would return the more familiar:
\begin{equation*}
\label{logIncResp}
    \log(\tilde{\lambda}_{\boldsymbol{\gamma}}(t))=\log(r) + \log\left( \tilde{\lambda}_{\boldsymbol{\gamma},-r}(t)\right)
\end{equation*}
From Equation \ref{logIncResp}, it comes natural the idea of expressing $\log(r)$ at each time-point $t$ as the linear combination of covariates $\eta(\mathbf{x}(t))$ as in the classic GLM Poisson model with $\log$ link function.
This indeed provides a multiplicative effect of the covariates, where the parameter $r$ is:
\begin{equation}
\label{multCovs}
r_{\boldsymbol{\beta}}(t)=\exp\left\lbrace\mathbf{x}(t)\boldsymbol{\beta}\right\rbrace=\exp\left\lbrace\beta_0+\beta_1x_1(t)+\dots+\beta_kx_k(t)\right\rbrace.
\end{equation}
Note that the constant $r$ is still present and included in Equation \ref{multCovs} through the intercept $\beta_0$.
Therefore, the mean at time $t$ is expressed:
\begin{equation*}
        \tilde{\mu}_\theta(t) = \alpha + \exp\left\lbrace\mathbf{x}(t)\boldsymbol{\beta}\right\rbrace \cdot \tilde{\lambda}_{\boldsymbol{\gamma},-r}(t)
\end{equation*}
Considering the whole vector of observations, we would have the following vector of means:
\begin{equation*}
    \tilde{\boldsymbol{\mu}}_{\boldsymbol{\theta}} = \boldsymbol{\alpha} + \exp\left\lbrace\mathbf{X}\boldsymbol{\beta}\right\rbrace \cdot \tilde{\boldsymbol{\lambda}}_{\boldsymbol{\gamma},-r},
\end{equation*}
where $\boldsymbol{\alpha}=\alpha\cdot\mathbf{1}_T$.

\subsection{Model estimation}
\label{subsec:ModEst}

Parameters can be estimated by maximizing the log-likelihood $l(\boldsymbol{\theta}|\mathbf{y})$, where $\boldsymbol{\theta}$ in this case includes all the parameters the likelihood depends on (e.g. includes $\nu$ in the Negative Binomial case). This optimization problem does not have an analytical solution, and numerical maximization must  be used.  To improve computation, we derived analytical expressions for the gradient and Hessian of the two possible log-likelihoods (i.e. Poisson or Negative Binomial counts), making Fisher-scoring iteration very fast.
The expressions are reported in the Appendix.
Given the non-smooth shape of the objective function, we are at risk of being trapped by local maxima of the log-likelihood, depending on the initial conditions.
Therefore, in order to robustify the optimization procedure, a multistart procedure based on genetic algorithms has been used \citep{scru:13}.\\
Once an approximate point of maximum $\hat{\boldsymbol{\theta}}$ has been obtained, the inverse of the negative log-likelihood Hessian in $\hat{\boldsymbol{\theta}}$ (which corresponds to the \textit{Observed Fisher Information}) is used to approximate its standard errors and variance-covariance matrix according to the usual asymptotic properties:
\begin{equation*}
\hat{V}_{\boldsymbol{\theta}}=-\mathbf{H}\left(l(\hat{\boldsymbol{\theta}}|\mathbf{y})\right)^{-1},
\end{equation*}
where $\mathbf{H}$ denotes the Hessian matrix and the interval estimates for the parameters are directly derived through the asymptotic distribution $\hat{\boldsymbol{\theta}}\sim\mathcal{N}(\boldsymbol{\theta}, \hat{V}_{\boldsymbol{\theta}})$.
A similar theoretical result for predictions is not straightforward. Therefore, these are derived through a parametric double bootstrap procedure (\cite{efron2004estimation}, \cite{hall2006parametric}, \cite{efron2012bayesian}), which accounts for both the uncertainty from estimation and the randomness of the observations. In practice, re-sampled trajectories $\left\lbrace\mathbf{Y}_i\right\rbrace_{i=1}^B$ are obtained by simulating $B$ sets of parameters from their asymptotic distribution and computing $B$ mean functions trajectories $\left\lbrace\mu_{\boldsymbol{\theta}_i}(\mathbf{t})\right\rbrace_{i=1}^B$. An artificial time series of counts is then simulated for each of the $B$ trajectories and $95\%$ confidence intervals are obtained by computing the point-wise $2.5\%$ and $97.5\%$ quantiles.
Diagnostic check on the model has been performed through the \textit{Pearson residuals}, which can be computed as:
\begin{equation*}
    \hat{\rho_t}=\frac{y_t-\hat{y_t}}{\widehat{\text{Var}}\left[Y_t\right]},\qquad t=1,\dots,T.
\end{equation*}
Recall that, under the \textit{Poisson} assumption $\widehat{\text{Var}}_{\text{Poi}}\left[Y_t\right]=\hat{y}_t$, while under the \textit{Negative Binomial} assumption $\widehat{\text{Var}}_{\text{NB}}\left[Y_t\right]=\hat{y}_t+\hat{y}_t^2/\hat{\nu}$.
Under the null hypothesis (the GLM assumption holds), \textit{Pearson residuals} are expected to be \textit{Normally distributed} and \textit{independent}.

\section{Nowcasting the Italian outbreak of COVID-19}
\label{Sec:NCItOutbreak}

For the sake of brevity, here we present results referred to the proposed \textit{Richards' growth model} only for \textit{daily positives} and \textit{daily deceased}, aggregated at the national level.
We use the Negative Binomial distributional assumption. Indeed, this choice is justified by the substantial over-dispersion present at the national level for these indicators.

We first show the fitted curve for each indicator, and compare its shape with the observed time series. We also calculate the residuals and check if model assumptions under the GLM framework hold.\\
Later on, we show the performance for two fundamental issues: (i) predicting the epidemic trend in advance and (ii) predicting the \textit{peak} of the epidemic.

\subsection{Model on \textit{daily positives}}
\label{Sec:modNP}

To decide whether or not to include the kink effect, we fitted the model with and without the baseline $\alpha$ and compared the two fits in terms of \textit{log-likelihood}, \textit{AIC}, \textit{BIC} and \textit{Corrected AIC (AICc)}.
The values are presented in Table \ref{tab:NPllAic} and provide clear evidence in favor of the model with baseline (i.e. with mean $\mu_t(\cdot)$ as in Eq. \ref{eqRespFunBase}).
Parameters' estimate of such model $\boldsymbol{\hat{\theta}}$ and the respective $95\%$ confidence intervals are shown in Table \ref{tab:NPEstBandr}, where the baseline $\alpha$ is estimated to be $\hat{\alpha}=175.04$, with interval $(158.15, 193.73)$, which confirms that the baseline is estimated to be significantly different from $0$, and it should be included in the model.

\begin{figure}[H]
    \begin{subfigure}{.49\textwidth}
    \centering
    \includegraphics[width=0.9\linewidth]{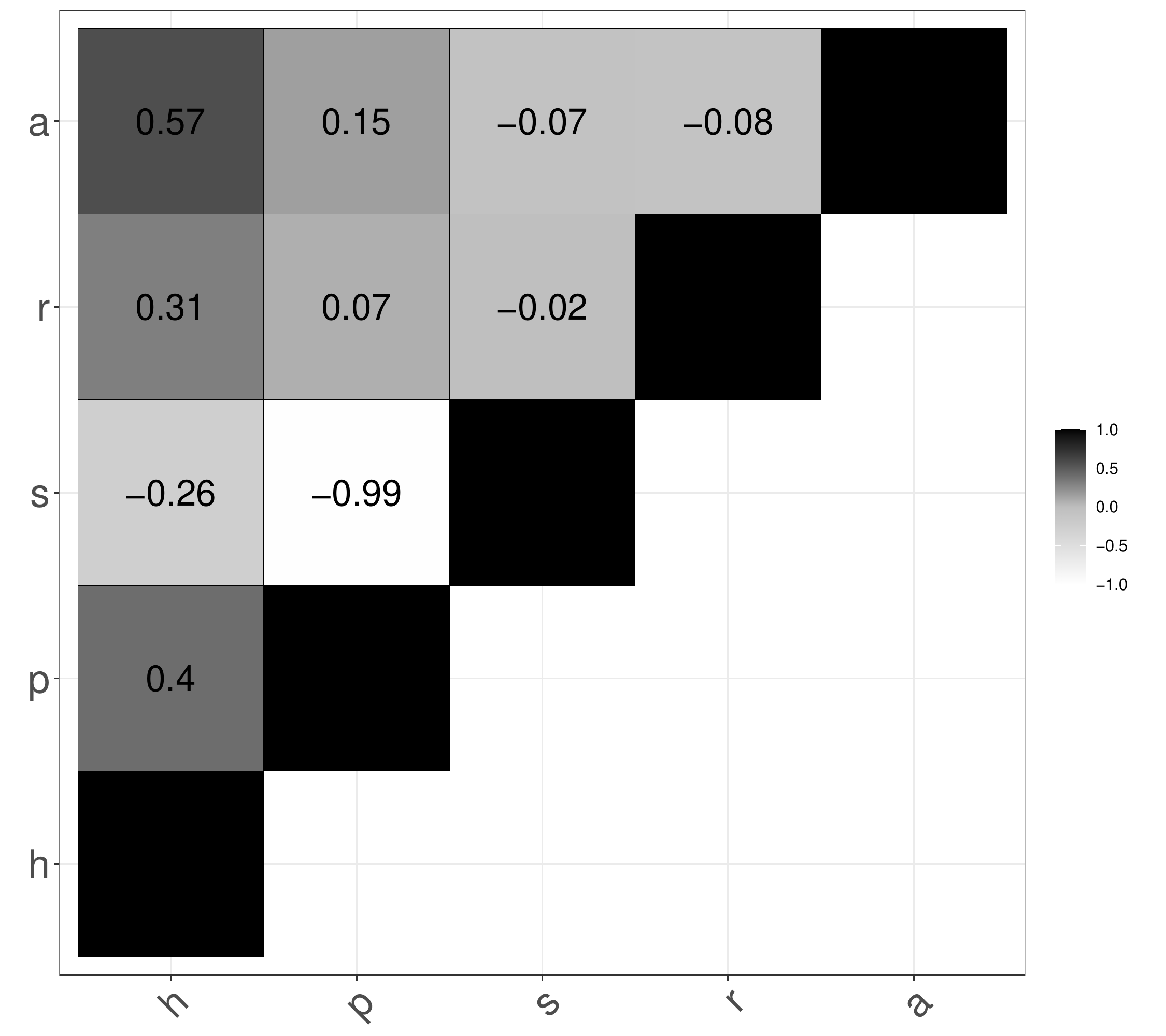}
    \caption{Correlation matrix}
    \label{fig:NPCorrparsBandr}
    \end{subfigure}
    \begin{subfigure}{.49\textwidth}
    \centering
    \includegraphics[width=0.9\linewidth]{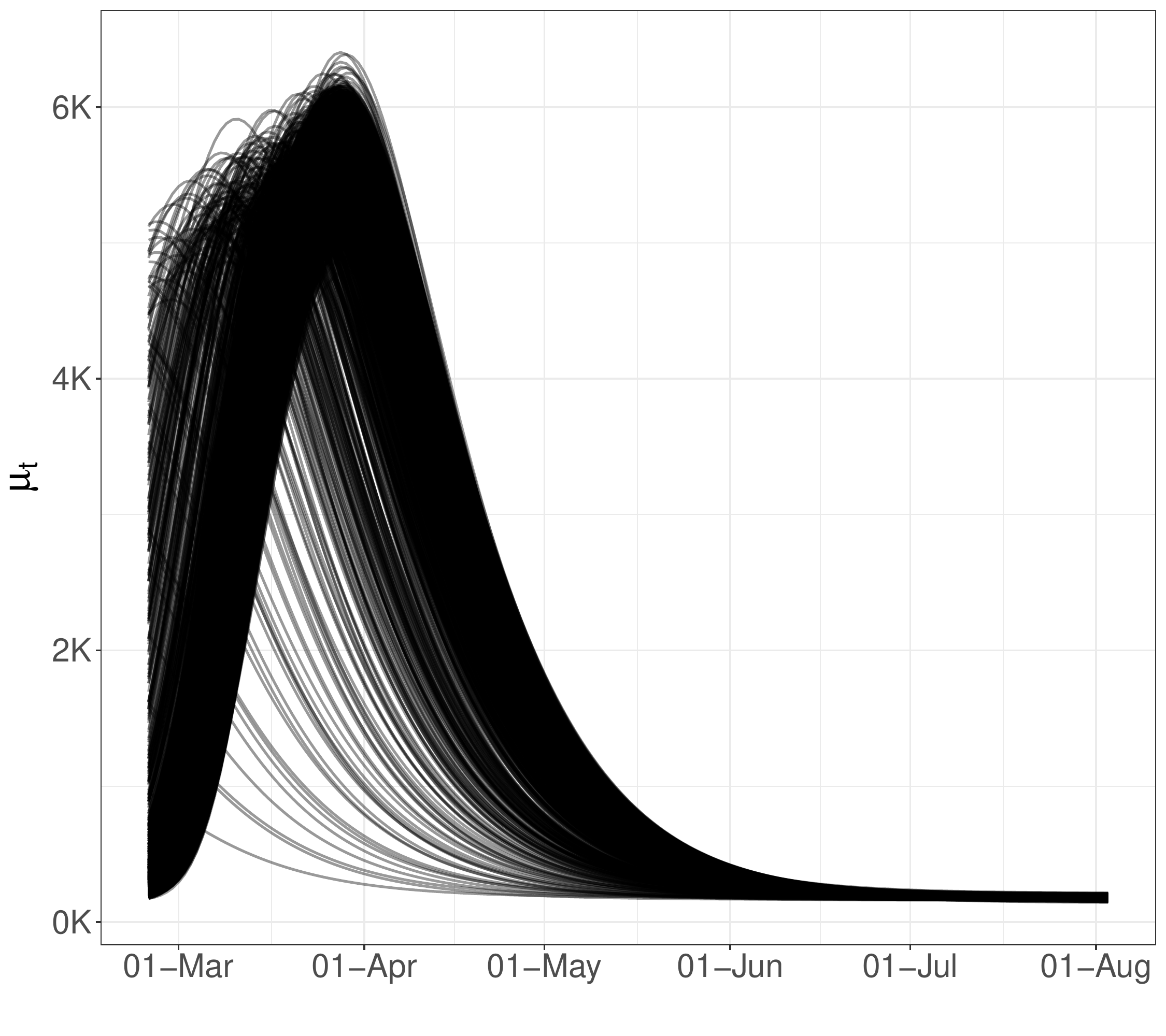}
    \caption{Trajectories}
    \label{fig:NPTrajsBandr}
    \end{subfigure}
    \caption{Correlation matrix and bootstrapped trajectories corresponding to the Hessian in the point of maximum for the model with baseline on \textit{daily positives}.}
    \label{fig:NPcorrproofBandr}
\end{figure}

The uncertainty characterizing the parameters $p$ and $s$ is not alarming. Indeed, according to the Hessian value in the maximum point, these two parameters are highly correlated (see Fig. \ref{fig:NPCorrparsBandr}).
Bootstrapping trajectories, by simulating $M=5000$ set of parameters from the Normal distribution with variance corresponding to the Hessian underlying the correlation matrix in Fig. \ref{fig:NPCorrparsBandr}, we get the set of curves in Fig. \ref{fig:NPTrajsBandr}.

\begin{table}[tp]
\centering
\caption{\textit{Log-likelihood, AIC, BIC and AICc for the model without baseline and the model with baseline, on daily positives}}
\begin{tabular}{lcc}
\toprule
\textbf{Index} & {Model without baseline} & {Model with baseline} \\
\midrule
\textit{log-likelihood} & $-1081.4$  & $-982.8$ \\
\textit{AIC} & $2152.7$  & $1953.6$ \\
\textit{BIC} & $2162.3$  & $1965$ \\
\textit{AICc} & $2137.8$  & $1935.7$ \\
\bottomrule
\end{tabular}
\label{tab:NPllAic}
\end{table}

\begin{table}[tp]
\centering
\caption{\textit{Parameters' points estimates and $95\%$ confidence intervals for the model with baseline, on daily Positives}}
\begin{tabular}{ccc}
\toprule
\textbf{Parameter} & {Point estimate} & {Interval}             \\
\midrule
$\alpha$    & $175.04$  & $(158.15, 193.73)$\\
$r$ & $221.94\times 10^3$& $(209.03\times 10^3, 235.65\times 10^3)$ \\
$h$ & $0.029$ & $(0.028, 0.03)$ \\
$p$ & $-32.29$ & $(-51.34, -13.23)$ \\
$s$ & $77.74$ & $(0, 170.88)$      \\
$\nu$ & $18.76$ & $(14.78, 23.81)$ \\
\bottomrule
\end{tabular}
\label{tab:NPEstBandr}
\end{table}

\begin{figure}[t]
    \centering
    \includegraphics[scale=0.4]{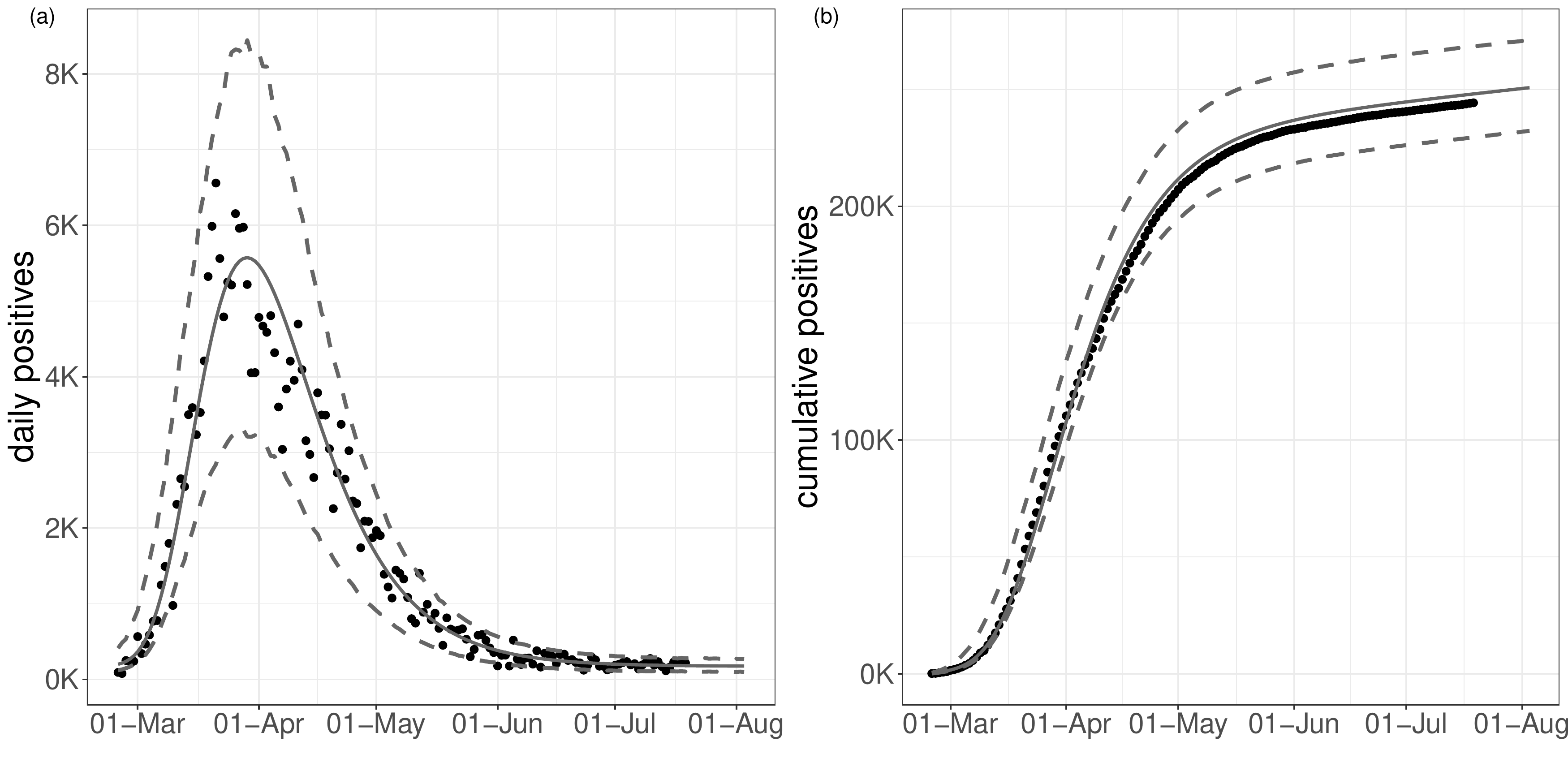}
     \caption{Observed (black dots) and fitted values (grey solid lines) with $95\%$ confidence intervals (grey dashed lines) for the model with baseline on \textit{daily positives}.}
     \label{fig:NPallFitBandr}
\end{figure}

We can also directly obtain point predictions $\left\lbrace \hat{y}_t\right\rbrace_{t=1}^T$ as:
\begin{equation}
    \label{EqPointPreds}
    \hat{y}_t=\mu_{\boldsymbol{\hat{\theta}}}(t), \quad t=1,\dots,T,
\end{equation}
and prediction intervals $\left\lbrace (\widehat{y}^l_t; \widehat{y}^u_t)\right\rbrace_{t=1}^T$ through the  same set of bootstrapped trajectories, whose statistical validity relies on the asymptotic properties introduced in Sec. \ref{subsec:ModEst}.
Fig. \ref{fig:NPallFitBandr} shows the model fit on the whole available time series of counts: the former on the daily series, the latter on the cumulative one. We can see how the estimated curve does catch the observed general behavior, providing a smooth approximation only marginally influenced by extreme values.
Fitting performances are further evaluated through numerical metrics such as the pseudo-$\text{R}^2$ and coverage of the $95\%$ prediction intervals:
\begin{equation*}
\begin{aligned}
    &\text{R}^2=1-\frac{\text{MSE}}{\sigma^2_y}=1-\frac{\sum_{t=1}^T(y_t-\hat{y}_t)^2}{\sum_{t=1}^T(y_t-\bar{y})^2},\\
    &\overline{\text{Cov}}_{95\%}=\frac{1}{T}\cdot\sum_{t=1}^T\mathbb{I}_{\left(\widehat{y}^l_t; \widehat{y}^u_t\right)}(y_t),
\end{aligned}
\end{equation*}
where $\text{MSE}$ is the \textit{Mean Squared Error}, $\bar{y}=\frac{1}{T}\sum_{t=1}^Ty_t$ and $\mathbb{I}_{\mathcal{Y}}(\cdot)$ denotes the indicator function over the set $\mathcal{Y}$.
Our model produces an $\text{R}^2=0.941$ and coverage $\overline{\text{Cov}}_{95\%}=0.945$, meaning that the percentage of observed daily counts falling inside the estimated bounds is perfectly coherent with the specified confidence level.
Looking at Fig. \ref{fig:NPallFitBandr}, we notice how daily counts boundaries get smaller as time passes, due to the implicit relationship between mean and variance that characterizes \textit{count} distributions. At the same time, the opposite happens to the bounds on the cumulative counts. The latter is not surprising: indeed, they are built marginally on all the epidemic's possible scenarios. Therefore, they give us a clear sight of what we could have currently observed, keeping into account and accumulating the uncertainty at each stage of the epidemic.

We also perform a diagnostic check on the Pearson residuals. In Fig. \ref{fig:NPResBandr}, we can observe different plots referred to the residuals: histogram (a), including the p-value from the Shapiro test;  Normal qq-plot (b);  autocorrelation plot (c); plot of the residuals vs. fitted values (d).
The first two check the (approximated) Normality assumption on the residuals, while the second two control for the correlation of the residuals (among them and with the observed values). %\vspace{-2.5cm}

\subsubsection{Weekly seasonality} %\vspace{-1cm}
\begin{figure}[t]
\centering
    \includegraphics[width=0.8\textwidth]{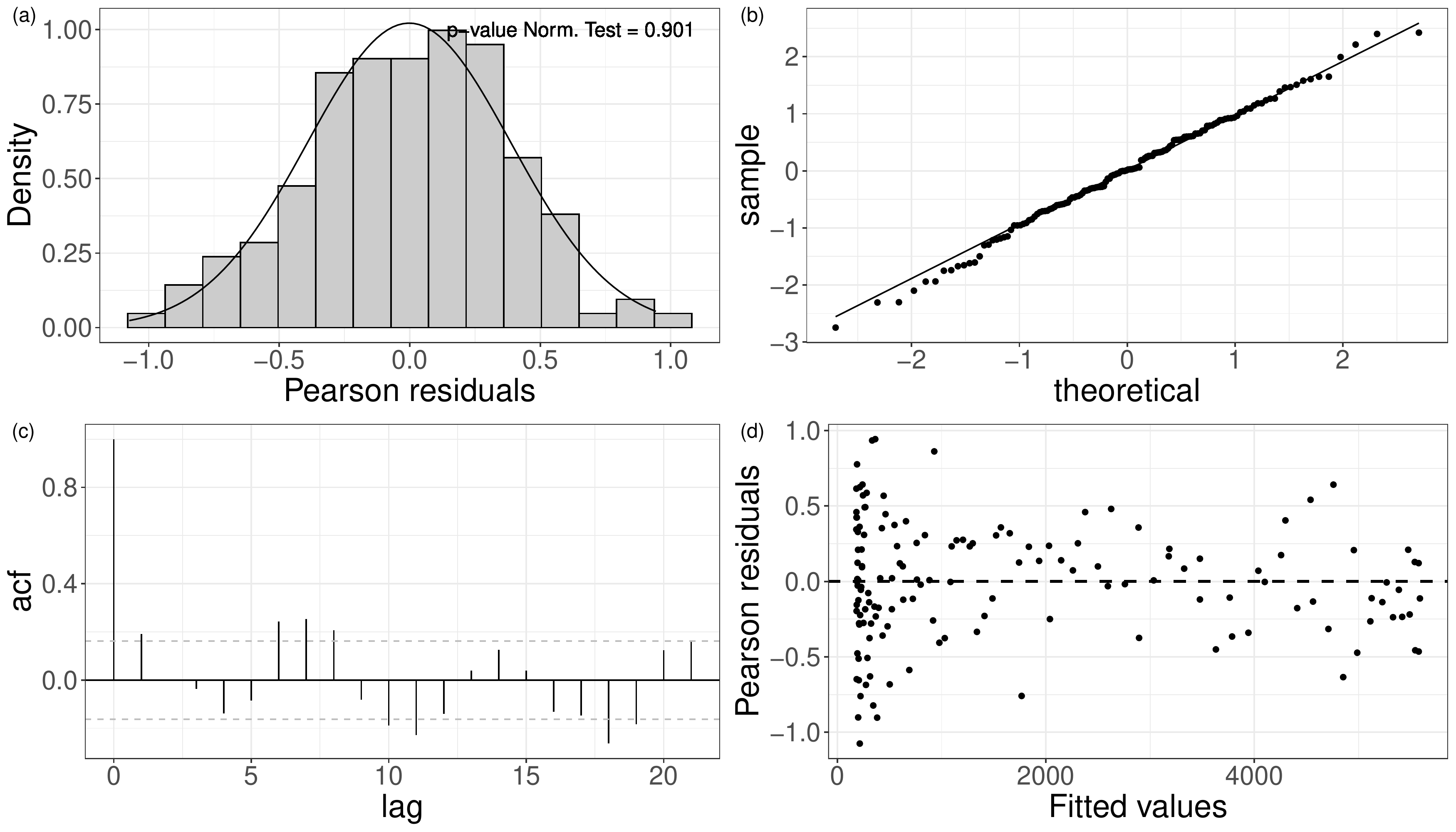}
    \caption{Pearson's residuals for the model with baseline on \textit{daily positives}.}
    \label{fig:NPResBandr}
\end{figure}

\begin{figure}[t]
    \centering\includegraphics[width=0.6\textwidth]{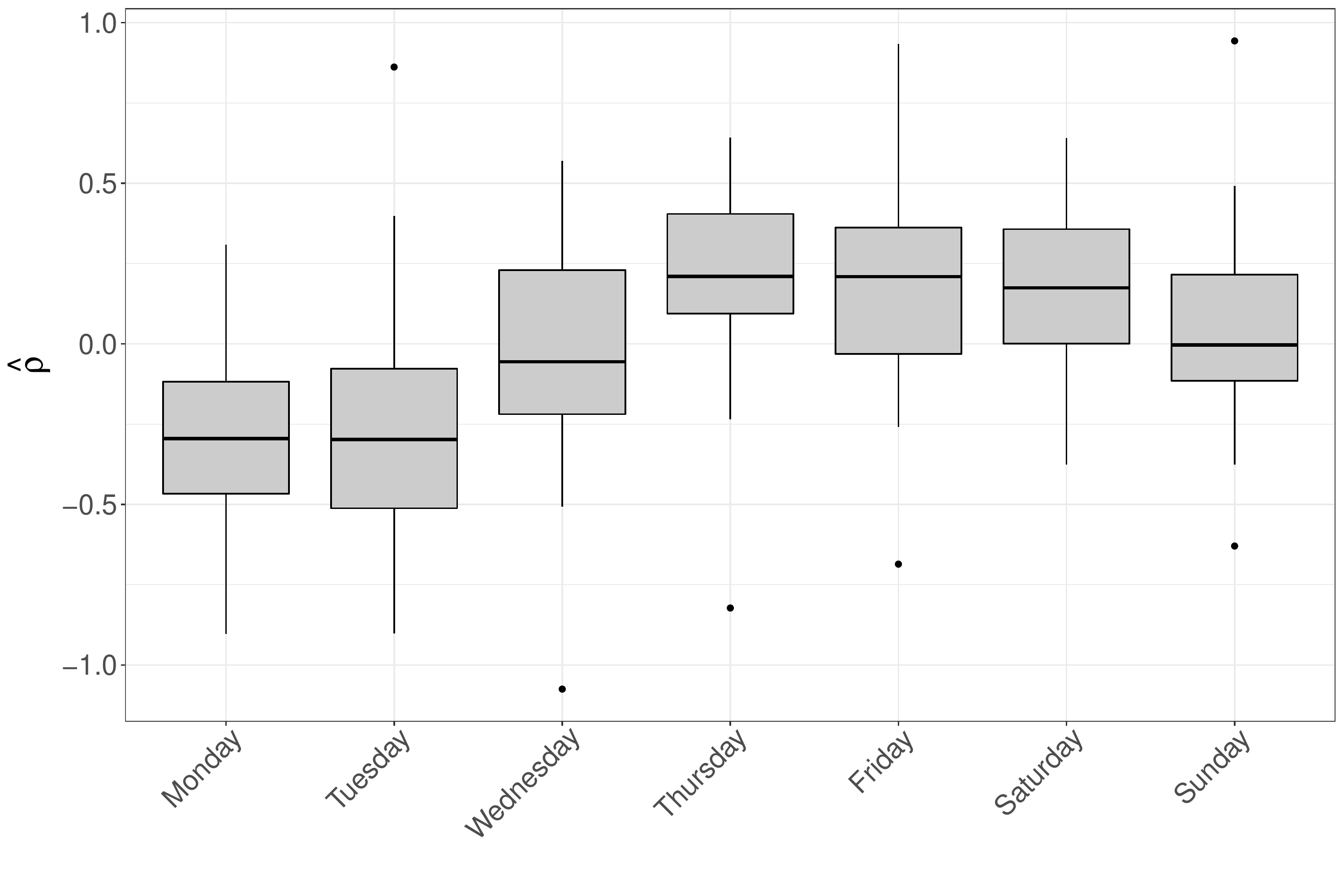}
    \caption{Pearson's residuals distribution aggregated by day of the week for \textit{daily positives}.}
    \label{fig:NPBPPearsonWeek}
\end{figure}

The diagnostic check on the residuals (see Fig. \ref{fig:NPResBandr}) shows that the Normality assumption is not rejected, but the correlation plot manifests undesirable patterns. In particular, the autocorrelation between errors is larger at lag $7$ (and multiples of this). We can interpret this outcome as the presence of an intense weekly seasonality (especially during/after the weekend). This may be adjusted by simply adding a weekday effect in our model as a covariate, using the approach in Sec. \ref{Subsec:IncCovs}. Such effect may be included either in an additive or a multiplicative fashion.
At first we considered effects for each day of the week, taking \textit{Monday} as a corner point. Preliminary results showed that not all week-days present a significant deviation from the common mean. On the other hand, the distribution of the Pearson residuals $\hat{\rho}_t$ of the standard model aggregated by week-day (see Fig. \ref{fig:NPBPPearsonWeek}) shows that an evident overestimation pattern (i.e., negative deviations) is taking place on Monday and Tuesday.
 
\begin{table}[t]
\centering
\caption{\textit{Log-likelihood, AIC, BIC and AICc for the models with baseline including additive or multiplicative week-day effect on daily positives}}
\begin{tabular}{lcc}
\toprule
\textbf{Index} & {Additive effect} & {Multiplicative effect} \\
\midrule
\textit{log-likelihood} & $-971.8$  & $-974.1$ \\
\textit{AIC} & $1929.6$  & $1934.3$ \\
\textit{AICc} & $1942.8$  & $1947.5$ \\
\textit{BIC} & $1908.7$  & $1913.4$ \\
\bottomrule
\end{tabular}
\label{tab:NPllAic_week}
\end{table}

Therefore, in the sequel, we will present only results obtained with the dichotomous variable that is equal to 1 whenever the week-day is Monday or Tuesday (0 vice versa). Note that lower tests effort during the weekend shows in the data on Monday and Tuesday, since daily reports involve mostly results received the day before, with swabs therefore dating back 48 hours on the day of publication.
The additive option is chosen over its alternative because of its lower/improved \textit{AIC}, \textit{BIC} and \textit{AICc} score (see Table \ref{tab:NPllAic_week}).

\begin{figure}[t]
    \centering
    \includegraphics[scale=0.4]{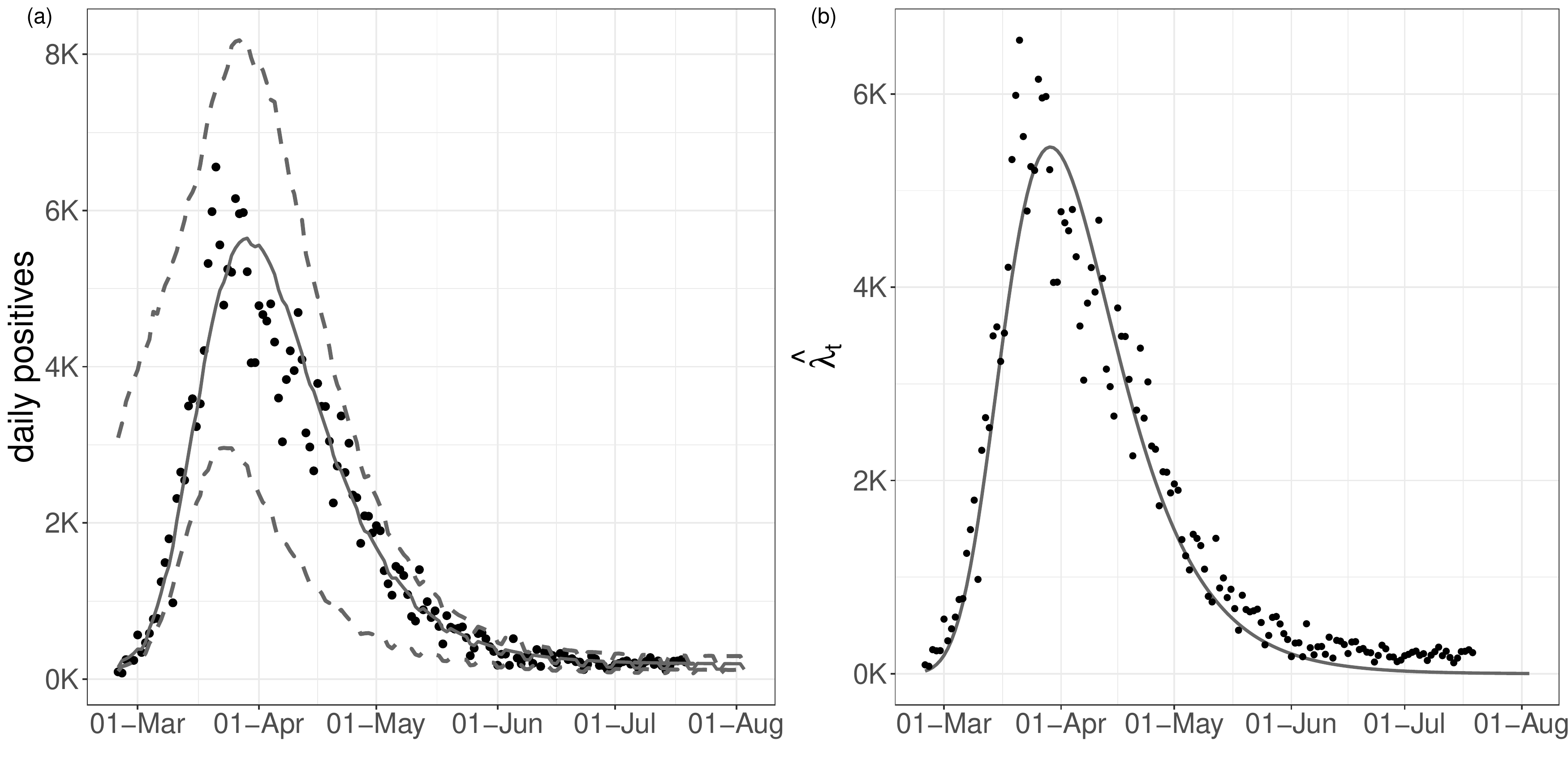}
     \caption{Fitted curve and $95\%$ prediction intervals (on the left) and estimated underlying trend of the epidemic (on the right), for the model with baseline and week-day additive effect, estimated on the \textit{daily positives}.}
     \label{fig:NPFitBandrWeek}
\end{figure}

The resulting fit of the model with week seasonality on the observed data is shown in Fig. \ref{fig:NPFitBandrWeek}, where, on the left, we show the fitted curve and the $95\%$ confidence intervals; on the right, we can show the first differences of the corresponding Richard's curve, not including the multiplicative effect of the week-days coefficients (the latter can be interpreted as the underlying trend of the epidemic, disentangled from the heterogeneity due to the week-days correlation). Estimated parameters are shown in Table \ref{tab:NPEstBandrWeek}.

\begin{table}[t]
\centering
\caption{\textit{Intercept $\beta_0$ and week-day effect $\beta_{wd}$ point estimates and $95\%$ confidence intervals for the additive model with baseline on daily positives}}
\begin{tabular}{ccc}
\toprule
\textbf{Parameter} & {Point estimate} & {Interval}             \\
\midrule
$\beta_{0}$  & $5.26$ & $(5.16, 5.36)$ \\
$\beta_{wd}$  & $-0.46$ & $(-0.63, -0.28)$                       \\
\bottomrule
\end{tabular}
\label{tab:NPEstBandrWeek}
\end{table}

\begin{figure}[H]
\centering
    \includegraphics[width=0.8\textwidth]{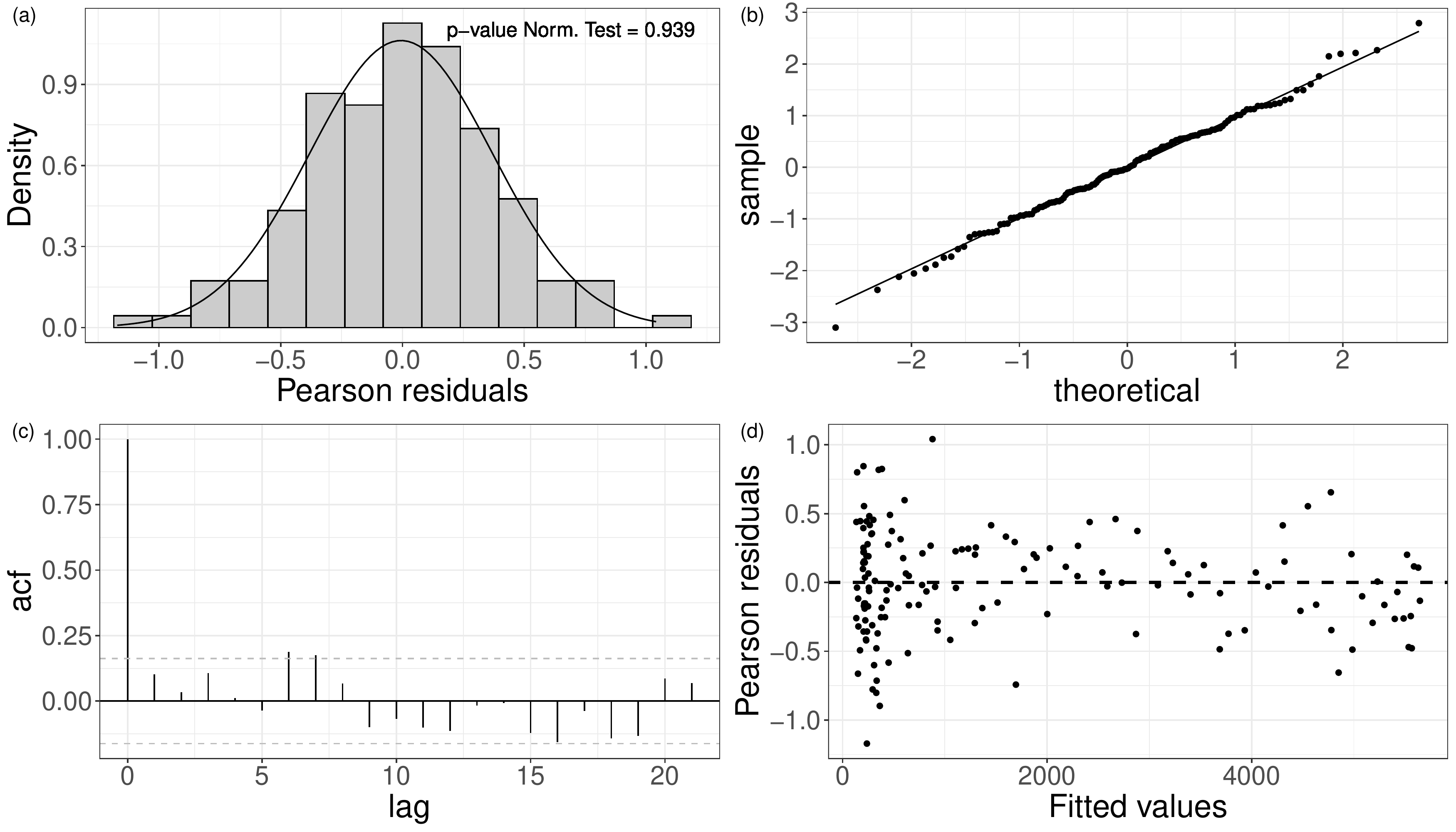}
    \caption{Pearson's residuals for the model with baseline and week-day additive effect estimated on \textit{daily positives}.}
    \label{fig:NPResBandrWeek}
\end{figure}

The inclusion of this effect improves sensibly the $R^2$ ($0.956$), while the average coverage $\overline{\text{Cov}}_{95\%}$ is constant ($0.950$).
Furthermore, the diagnostic check of the Pearsons's residuals (see Fig. \ref{fig:NPResBandrWeek}) shows that the correlation pattern at lag $7$ has diminished.

\subsection{Model on \textit{daily deceased}}
\label{Sec:ModDt}

\begin{table}[t]
\centering
\caption{\textit{Log-likelihood, AIC, BIC and AICc for the model without baseline and the model with baseline on daily deceased}}
\begin{tabular}{lcc}
\toprule
\textbf{Index} & {Model without baseline} & {Model with baseline} \\
\midrule
\textit{log-likelihood} & $-735.8$  & $-732.1$ \\
\textit{AIC} & $1461.6$  & $1452.11$ \\
\textit{AICc} & $1471.2$  & $1463.5$ \\
\textit{BIC} & $1446.7$  & $1434.2$ \\
\bottomrule
\end{tabular}
\label{tab:DtllAic}
\end{table}

\begin{table}[t]
\centering
\caption{\textit{Parameters' points estimates and $95\%$ confidence intervals for the model with baseline on daily deceased}}
\begin{tabular}{lcc}
\toprule
\textbf{Parameter} & Point estimate & Interval          \\
\midrule
$\alpha$ & $3.74$ & $(1.74, 8.05)$ \\
$r$   & $35.42\times 10^3$   & $(33.13\times 10^3, 37.87\times 10^3)$ \\
$h$  & $0.025$  & $(0.023, 0.026)$  \\
$p$   & $-50.4$  & $(-63.7, -37.04)$ \\
$s$     & $170.6$  & $(50, 291.2)$ \\
$\nu$    & $11.9$    & $(8.82, 16.18)$ \\
\bottomrule
\end{tabular}
\label{tab:DtEstBandr}
\end{table}

The procedure described in Sec. \ref{Sec:modNP} has been applied to decide for the inclusion of the baseline in the modeling effort of the \textit{daily deceased}, too. Comparisons in terms of goodness of fit measures are reported for both models in Table \ref{tab:DtllAic}.
The best model in terms of all the goodness of fit scores (AIC, AICc and BIC) is the model with baseline. The resulting estimated parameters $\boldsymbol{\hat{\theta}}$ and the respective intervals are shown in Table \ref{tab:DtEstBandr}, where the baseline $\alpha$ is estimated to be $\hat{\alpha}=3.74$ (sensibly larger than $0$).

\begin{figure}[t]
    \centering
    \includegraphics[scale=0.4]{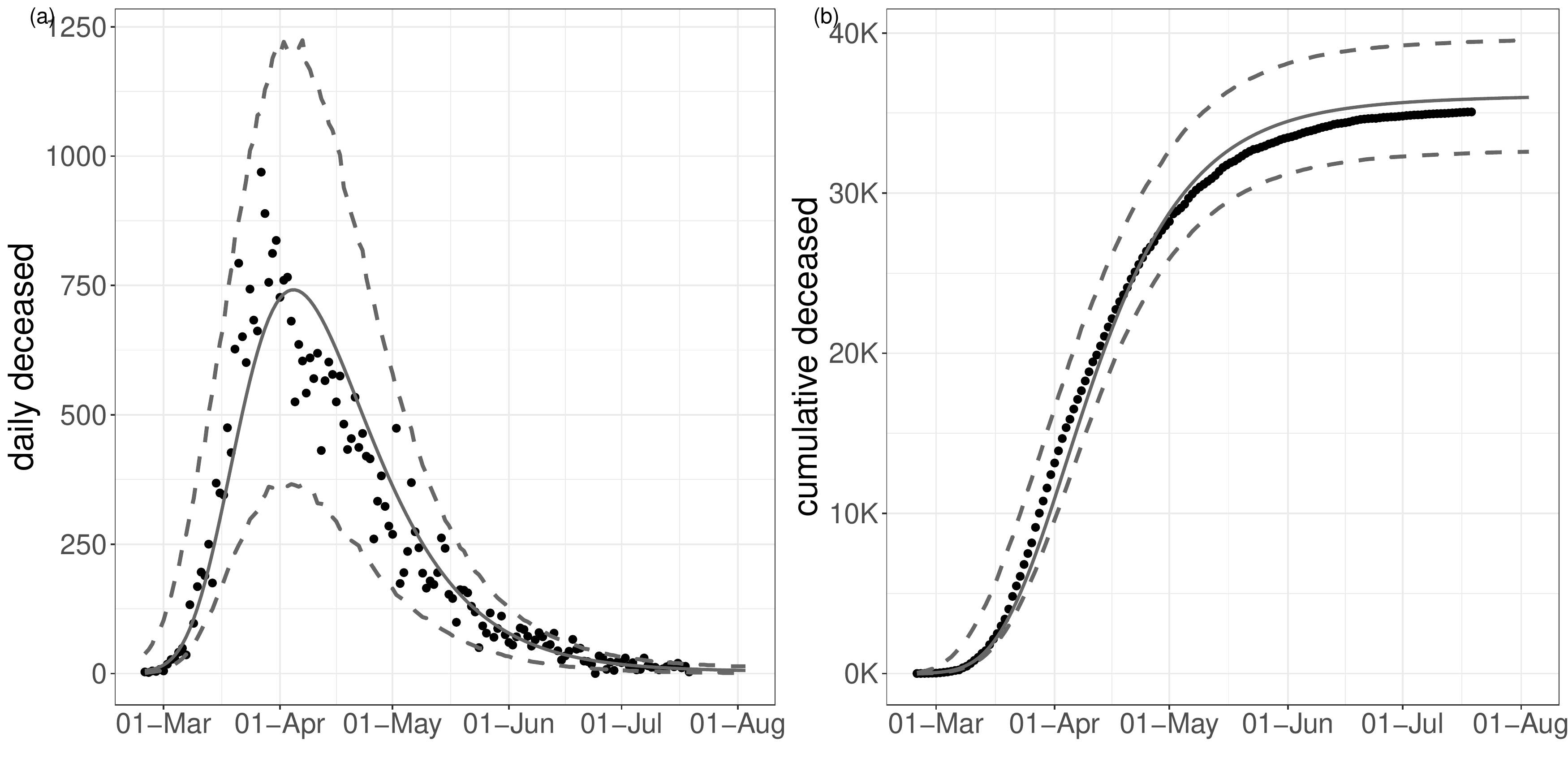}
     \caption{Observed (black dots) and fitted values (grey solid lines) with $95\%$ confidence intervals (grey dashed lines) for model with baseline on \textit{daily deceased}.}
     \label{fig:DtallFitBandr}
\end{figure}

\begin{figure}[t]
\centering
    \includegraphics[width=0.8\textwidth]{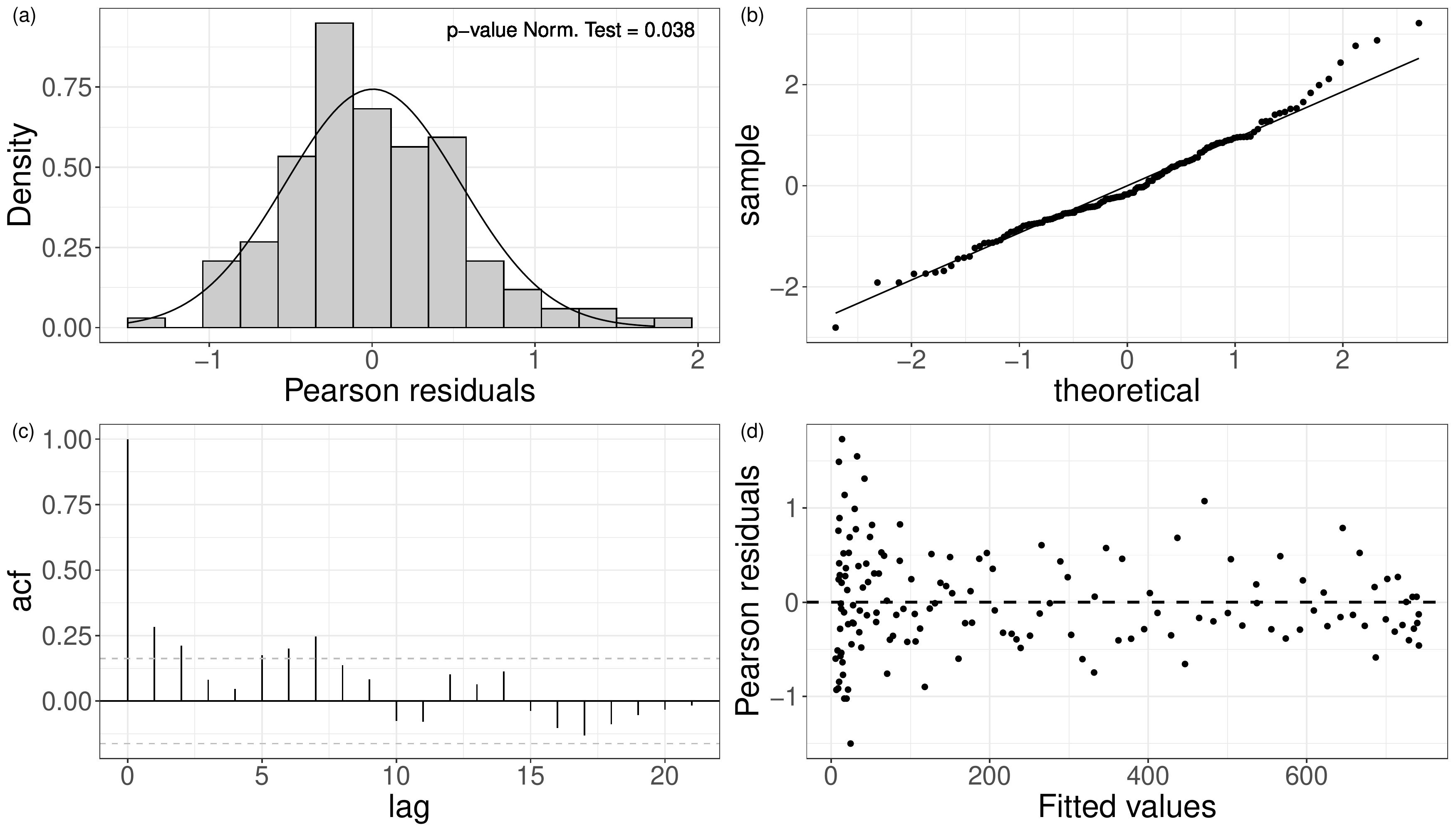}
    \caption{Pearson's residuals for the model with baseline on \textit{daily deceased}.}
    \label{fig:DtResBandr}
\end{figure}

We can then obtain point predictions $\left\lbrace \hat{y}_t\right\rbrace_{t=1}^T$ through Eq. \ref{EqPointPreds} and prediction intervals $\left\lbrace (\widehat{y}^l_t; \widehat{y}^u_t)\right\rbrace_{t=1}^T$ through the parametric bootstrap procedure described in Sec. \ref{subsec:ModEst}.\\
Fig. \ref{fig:DtallFitBandr} shows the fit on the whole available time series of counts: the former on the daily series, the latter on the cumulative one. Also in the case of the deceased the estimated curve does catch the observed general behavior.
The same metrics are used to evaluate the fitting performances, which correspond to an $R^2=0.90$ and a coverage $\overline{\text{Cov}}_{95\%}=0.95$.
Pearson residuals are shown in Fig. \ref{fig:DtResBandr}.

\begin{table}[H]
\centering
\caption{\textit{Log-likelihood, AIC, BIC and AICc for the models with baseline including additive or multiplicative week-day effect on daily deceased}}

\begin{tabular}{lcc}
\toprule
\textbf{Index} & {Additive effect} & {Multiplicative effect} \\
\midrule
\textit{log-likelihood} & $-725.71$  & $-725.30$ \\
\textit{AIC} & $1437.43$  & $1436.61$ \\
\textit{AICc} & $1450.62$  & $1449.81$ \\
\textit{BIC} & $1416.55$  & $1415.73$ \\
\bottomrule
\end{tabular}
\label{tab:DtllAic_week}
\end{table}

\begin{table}[H]
\centering
\caption{\textit{Intercept $\beta_0$ and week-day effect $\beta_{wd}$ point estimates and $95\%$ confidence intervals for the additive model with baseline on daily deceased}}
\begin{tabular}{ccc}
\toprule
\textbf{Parameter} & {Point estimate} & {Interval}             \\
\midrule
$\beta_{0}$  & $10.54$ & $(10.46, 10.61)$ \\
$\beta_{wd}$  & $-0.22$ & $(-0.33, -0.1)$                       \\
\bottomrule
\end{tabular}
\label{tab:DtEstBandrWeek}
\end{table}

\newpage

\begin{figure}[t]
    \centering\includegraphics[width=0.6\textwidth]{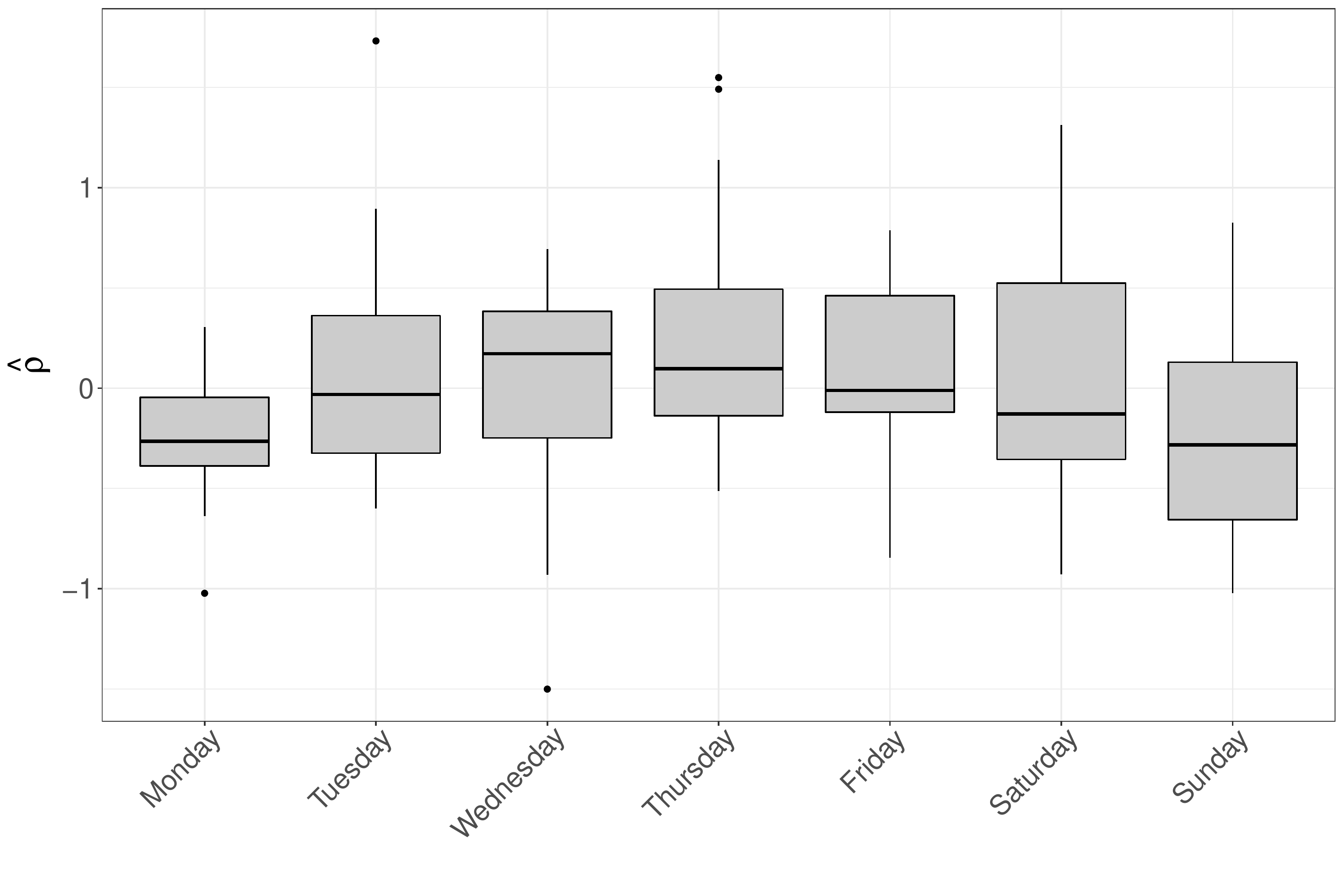}
    \caption{Pearson's residuals distribution aggregated by day of the week for \textit{daily deceased}.}
    \label{fig:DtBPPearsonWeek}
\end{figure}

\subsubsection{Week seasonality}
\begin{figure}[H]
    \centering
    \includegraphics[scale=0.4]{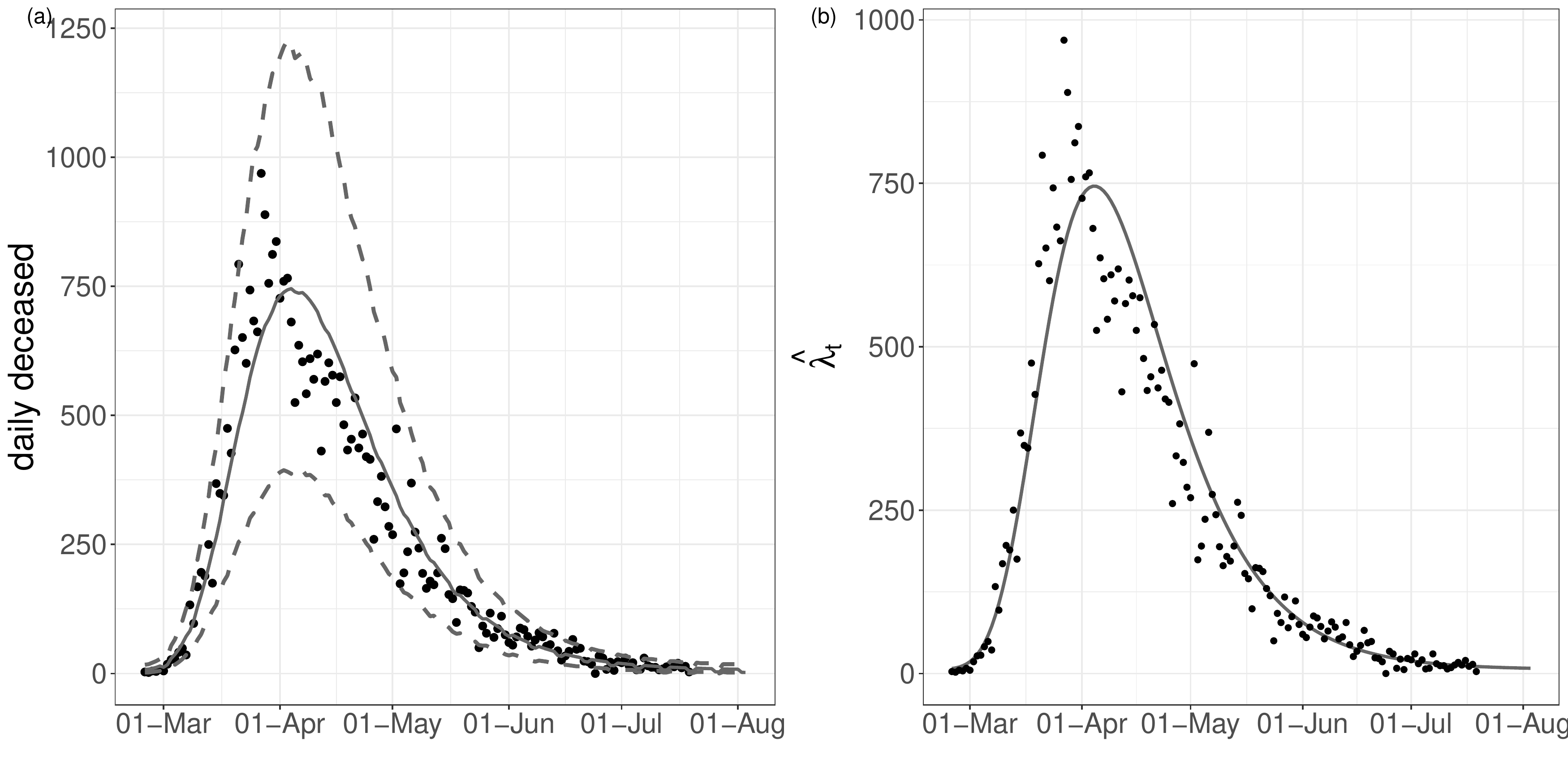}
    \caption{Fitted curve and $95\%$ prediction intervals (on the left) and estimated underlying trend of the epidemic (on the right), for the model with baseline and week-day additive effect, estimated on the \textit{daily deceased}.}
    \label{fig:DtFitBandrWeek}
\end{figure}

\begin{figure}[t]
\centering
    \includegraphics[width=0.8\textwidth]{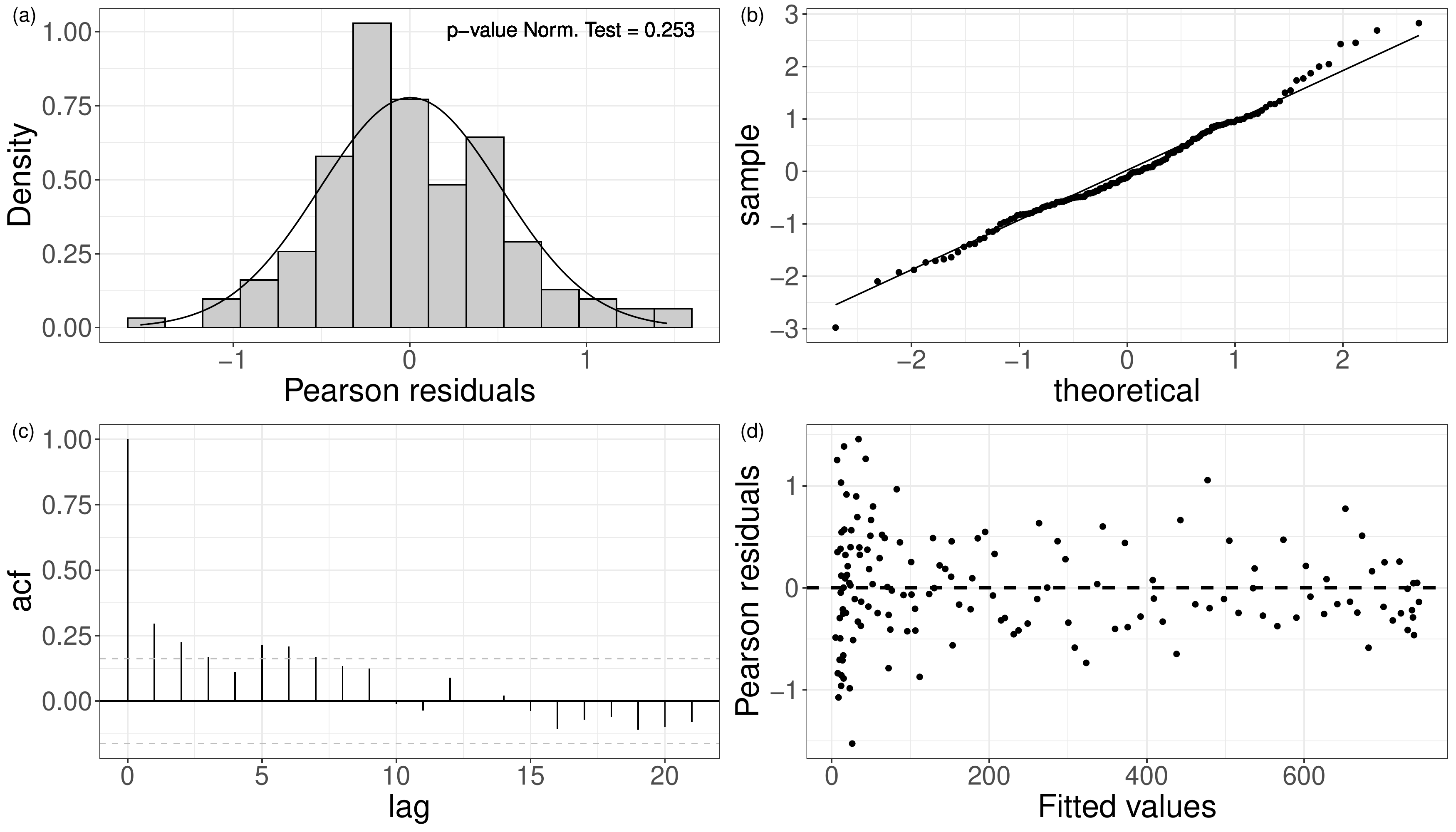}
    \caption{Pearson's residuals for the model with baseline and week-day additive effect on \textit{daily deceased}.}
    \label{fig:DtResBandrWeek}
\end{figure}

As in the case of \textit{daily positives}, the diagnostic check on the Pearson's residuals for the \textit{daily deceased} highlights a slight week seasonality pattern for the autocorrelations. In addition, also the residual Normality hypothesis is rejected. Potentially, the inclusion of a week-day effect may solve both problems. In order to decide what set of week-days to group together, we visualize the residuals' distribution aggregated by week (see Fig. \ref{fig:DtBPPearsonWeek}). The pattern is not as evident as in the case of \textit{daily positives}, but we can still detect some undesirable overestimation on Mondays and Sundays.
Therefore, on the line of the previous application, we decide to include a dichotomous week-day fixed effect on the pair Monday-Sunday. As before, this effect may be included either in an additive or a multiplicative fashion and, again, we may pick the version that achieves the best AIC, AICc and BIC scores. However, as shown in Table \ref{tab:DtllAic_week}, differences in this scores are almost negligible and choice based on such a small improvement would not be robust. Therefore, we checked the Pearson residuals for both alternatives and we selected the additive model because of the improved residuals behavior (Normality is accepted, autocorrelation at lag $7$ is reduced).
Estimated parameters are shown in Table \ref{tab:DtEstBandrWeek}, where the Monday-Sunday effect is estimated to have a reducing effect on the daily baseline rate of $\approx -357, i.e. \exp{\left\lbrace-357\right\rbrace}\approx 0$, which shrinks to $0$ the \textit{kink effect} on Mondays and Sundays. The resulting fit is shown in Fig. \ref{fig:DtFitBandrWeek} where: on the left, we can observe the fitted curve and the $95\%$ confidence intervals; on the right, we can observe the first differences of the corresponding Richard's curve, not including the multiplicative effect of the week-days coefficients (i.e. underlying trend of the epidemic).
The inclusion of the Sunday-Monday effect allows for an increase of the $R^2$ to $0.91$, whilst keeping the coverage $\overline{\text{Cov}}_{95\%}$ steady at $0.95$.
The diagnostic check shown in Fig. \ref{fig:DtResBandrWeek} shows how Residual Normality is now accepted and the previously evident correlation pattern is slightly reduced.

\newpage

\begin{figure}[t]
    \centering
     \includegraphics[width=0.8\textwidth]{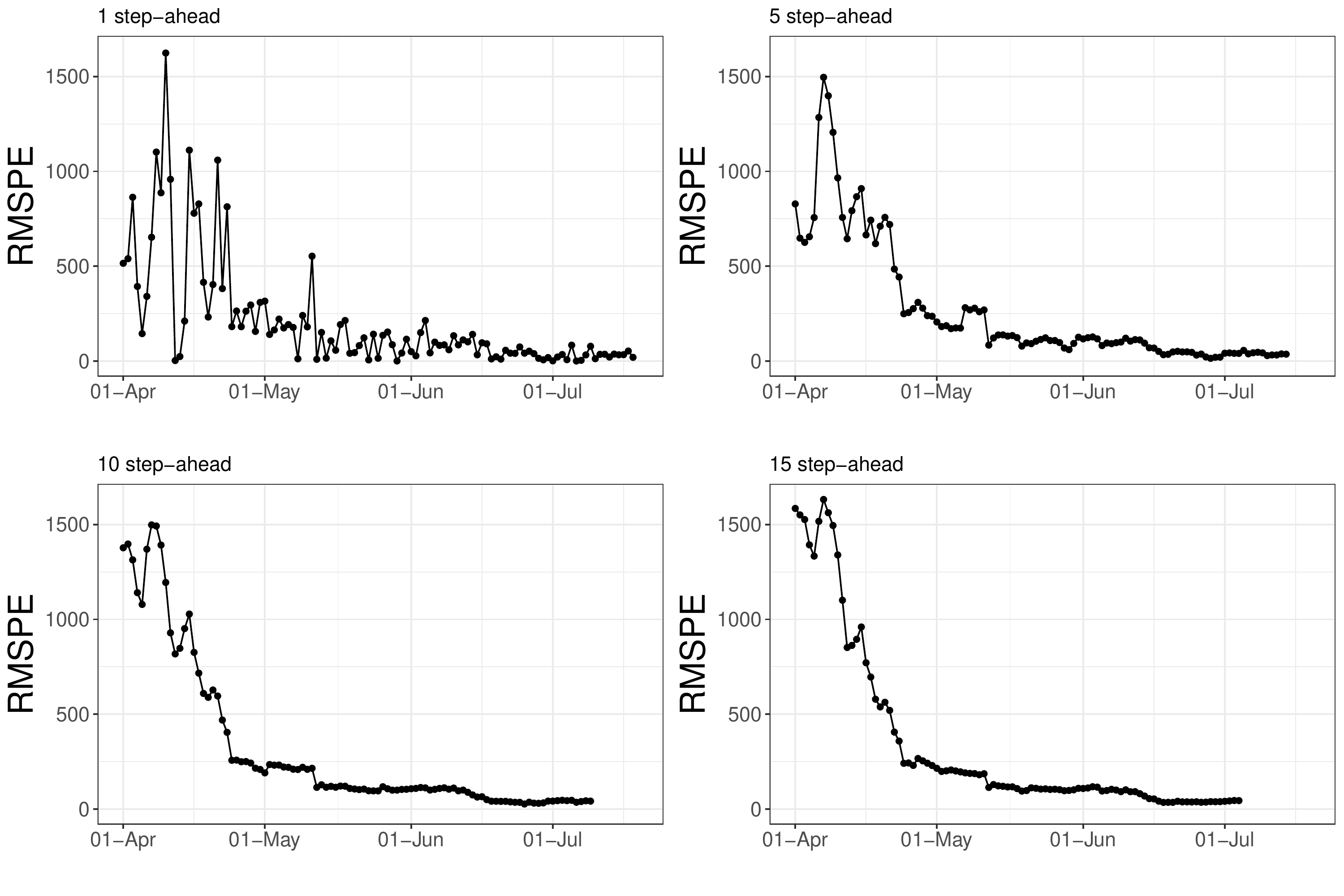}
    \caption{RMSPE for \textit{daily positives} at different steps-ahead.}
    \label{RMSENP}
\end{figure}

\subsection{Step-ahead predictions}
\label{Subsec:sahead}
In this section, we test our model's ability to predict the evolution of the epidemic (at least its first wave) from the short to the medium term.
Indeed, while the choice of a rigid parametric form for the mean function is penalizing in terms of flexibility and fitting ability, it allows for extrapolation outside the observed domain and is supposed to provide robust forecasts (at least in the short/medium term).
Therefore, using the best model for the two indicators (i.e. baseline + week-day additive effect), we calculated the \textit{out-of-sample} Root Mean Squared Prediction Error (RMSPE) for:
\begin{itemize}
    \item different fitting windows $t=1,\dots,\tilde{t}$, where $\tilde{t}$ goes from the $1$st of April up to the $19$th of July
    \item different forecast horizons, say $K \in \lbrace1, 5, 10, 15\rbrace$
\end{itemize}
We recall that, given the fitting window set $1,\dots, \tilde{t}$:
\begin{equation*}
    \text{RMSPE}_{\tilde{t},K} = \sqrt{\frac{1}{K}\sum_{j=1}^K(y_{\tilde{t}+j}-\hat{y}_{\tilde{t}+j})}
\end{equation*}

\begin{figure}[tp]
    \centering
    \includegraphics[width=0.8\textwidth]{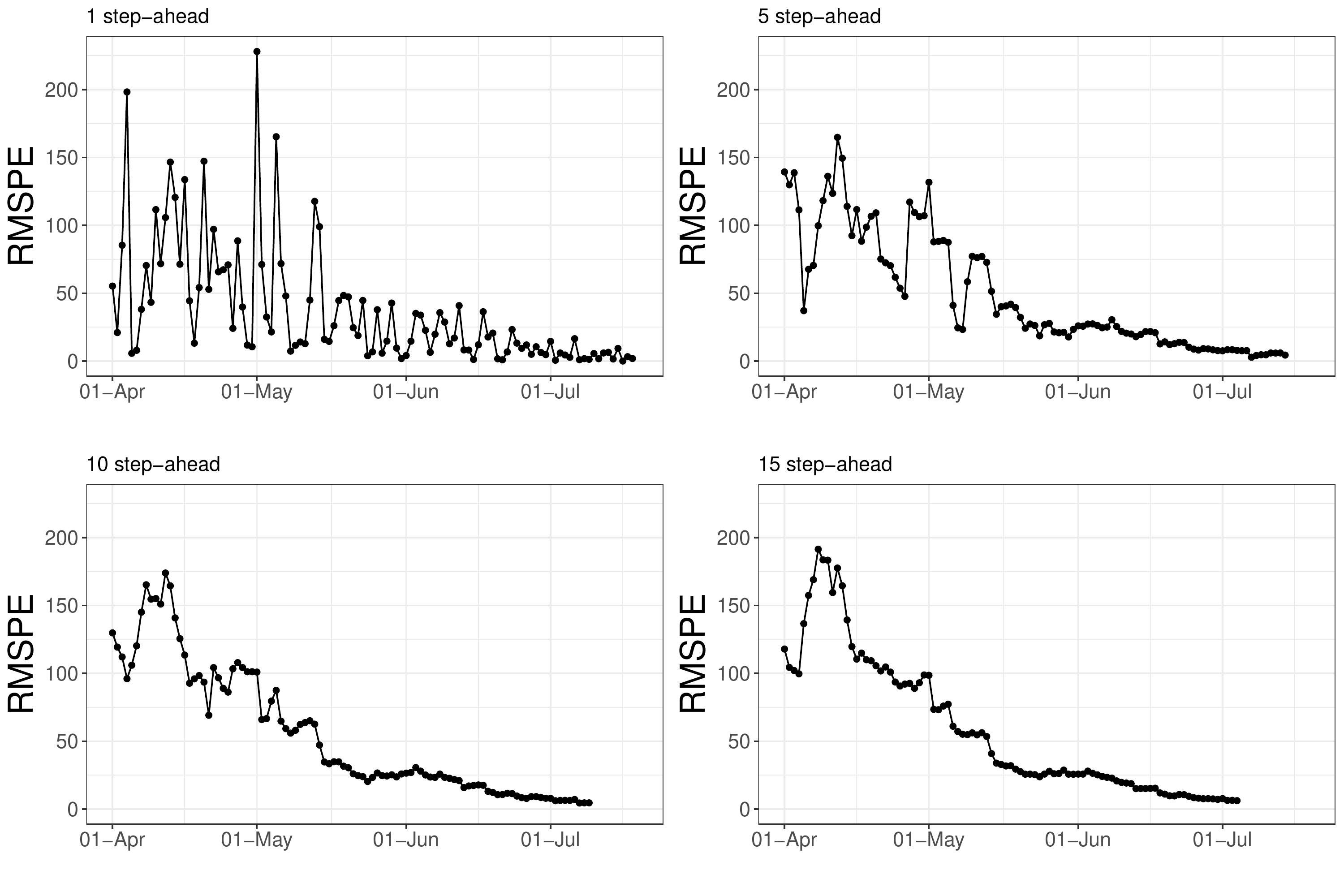}
    \caption{RMSPE for \textit{daily deceased} at different steps-ahead.}
    \label{RMSEDt}
\end{figure}

The RMSPEs for each steps-ahead are presented in Fig. \ref{RMSENP} and \ref{RMSEDt}. Results match the expectations as: (i) the error decreases with the length of the fitting window; (ii) the error trend is more stable on larger testing windows (10-15 steps ahead vs 1-5 steps ahead); (iii) larger errors are made around the day of the peak. It can be seen nevertheless that predictions are always reasonable at these time horizons.

\subsection{Prediction of the peak day}
\label{Subsec:pday}

\begin{figure}[tp]
    \centering\includegraphics[width=0.8\textwidth]{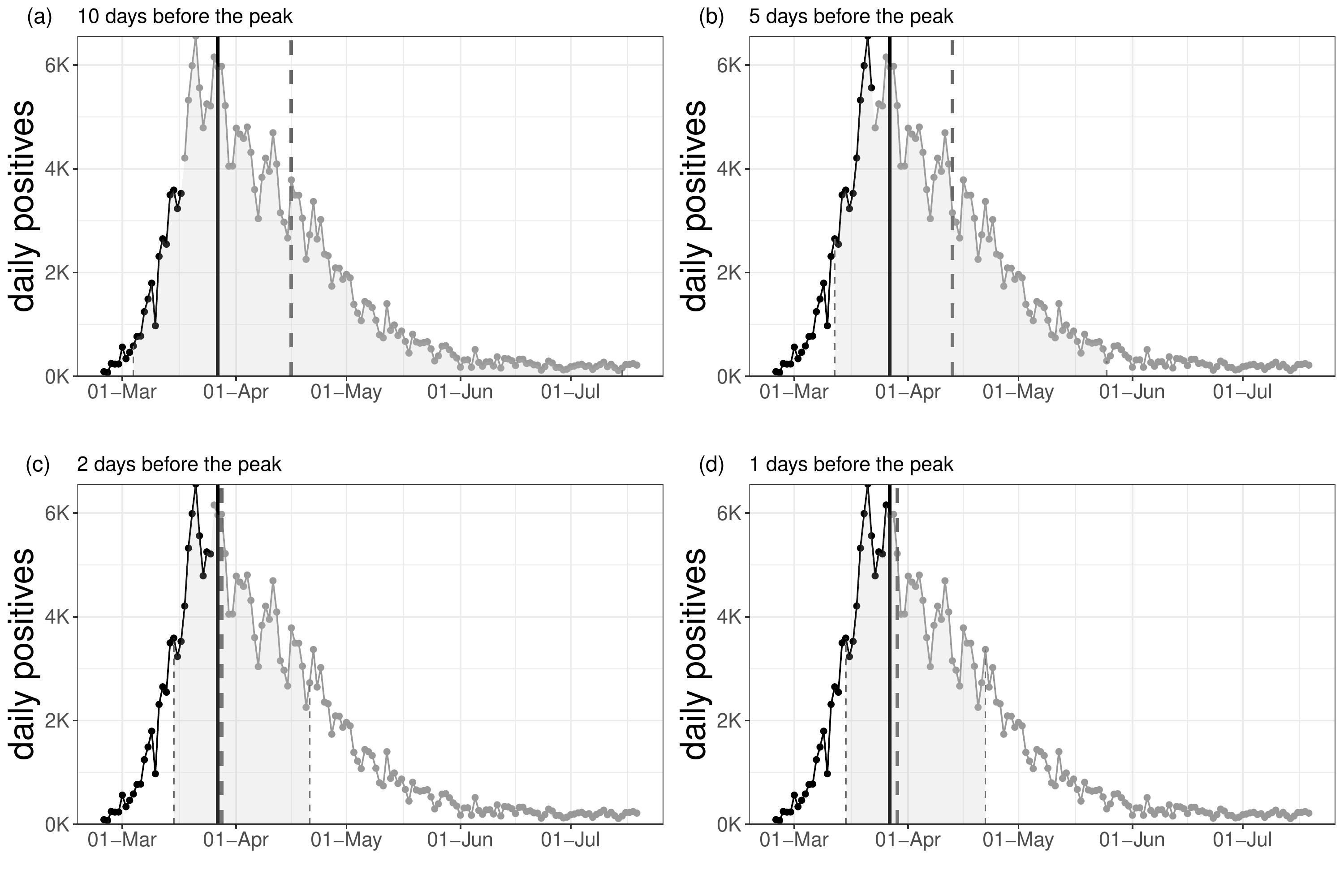}
    \caption{Estimation of the date of the peak for \textit{daily positives} at different steps-before.}
    \label{PeakEstNP}
\end{figure}

\begin{figure}[H]
    \centering\includegraphics[width=0.8\textwidth]{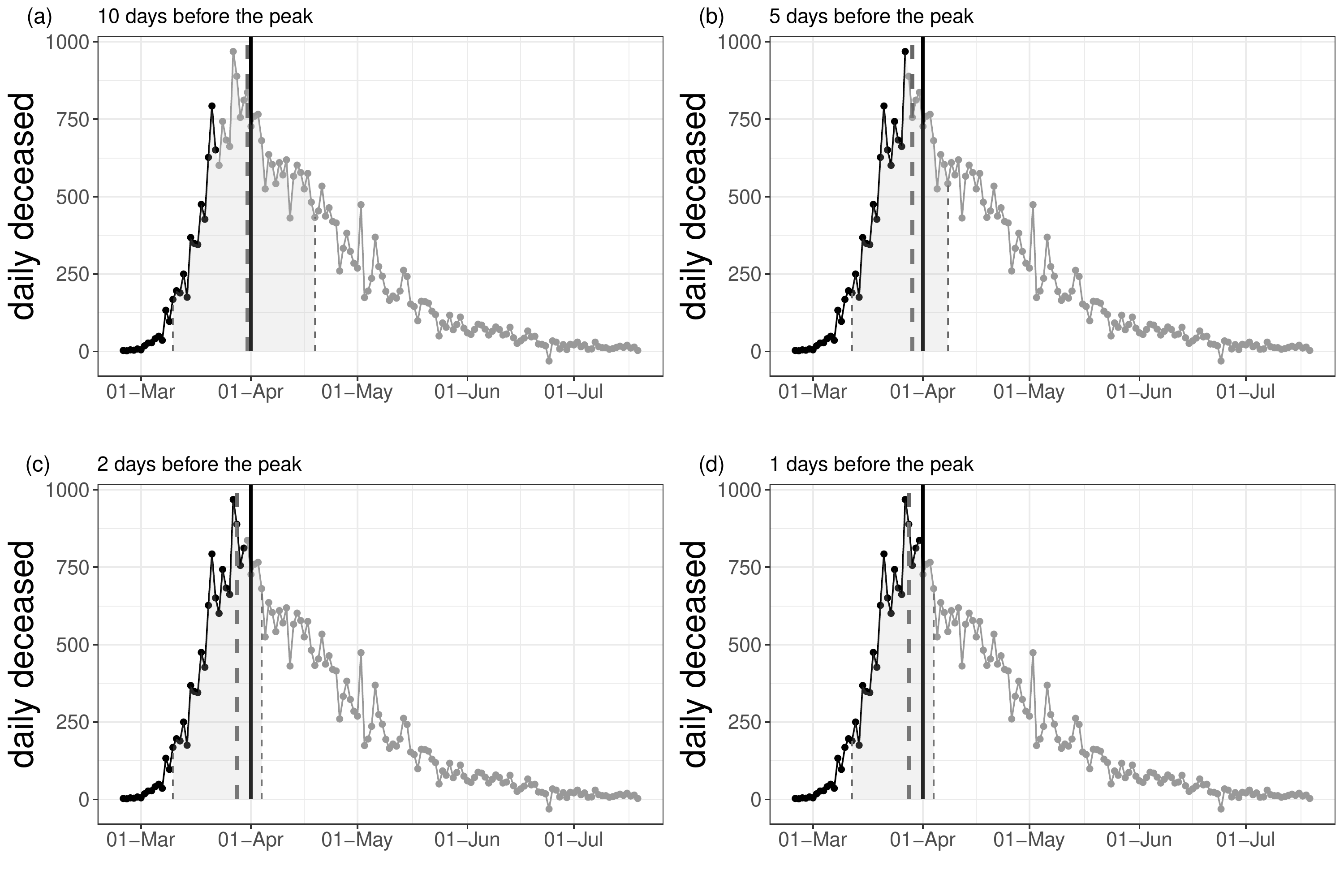}
    \caption{Estimation of the date of the peak for \textit{daily deceased} at different steps-before.}
    \label{PeakEstDt}
\end{figure}

Finally, we evaluate the model's ability to predict the date of the peak.
To do so, we estimate the model without covariates, using all  available data until $K \in \lbrace15, 10, 5, 3, 2, 1\rbrace$ days before the observed peak.
For the sake of conciseness, we only report results for $K \in \lbrace 10, 5, 2, 1\rbrace$ as shown in Fig. \ref{PeakEstNP} and \ref{PeakEstDt}.

When $s=1$, the peak $\hat{t}$ is directly expressed by the parameter $p$. When $s\neq 1$, after some algebra it can be seen that the peak can still be computed analytically as:
\begin{equation*}
    \hat{t}_{\boldsymbol{\gamma}} = \hat{p} + \frac{\log_{10}(\hat{s})}{\hat{h}}.
\end{equation*}

Confidence intervals are obtained through the same bootstrap procedure introduced in Sec. \ref{subsec:ModEst}.
The dashed grey vertical lines represent the bounds of the confidence interval and the predicted date of the peak (confidence area is shaded with the same grey). The solid vertical black line represents the \textit{"true"} date of the peak (i.e. obtained via smoothing of the observed counts through non-parametric polynomial approximations). The observed time-series is represented through point and lines, where the black section is referred to the training window while the grey section is referred to the testing (out-of-sample) window.

As expected, as we approach the real date of the peak, we predict it more accurately. Point predictions are very accurate for both indicators since 5 days before the actual peak. At the same time, interval bounds get tighter and tighter as the fitting interval approached the day of the peak and, in general, the day of the peak is always included in such bounds (see Table \ref{peakEstTab} for exact numerical evaluation).

\begin{table}[tp]
    \centering
    \caption{\textit{Delay (days) in point estimation of the peak}}
    \begin{tabular}{lcccccccc}
    \toprule
    \multirow{2}{*}{Days before}& \multicolumn{2}{c}{10} & \multicolumn{2}{c}{5}& \multicolumn{2}{c}{2} & \multicolumn{2}{c}{1} \\
     & Delay & Width & Delay & Width & Delay & Width & Delay & Width \\
        \midrule
        \textit{daily deceased} &       -1 & 37 & -3 & 25 & -4 & 22 & -3 & 21\\
        \textit{daily positives} &  20 & 106 & 17 & 69 & 1 & 37 &  2 & 37\\
        \bottomrule
    \end{tabular}
    \label{peakEstTab}
\end{table}

%\newpage
\section{Discussion and further work} 
\label{Sec:Duscussion}

We presented an approach to modeling and prediction of epidemic indicators that has proven useful during the first outbreak of COVID-19 in Italy. The model has been validated on publicly available data, and has proved flexible enough to adapt to different indicators. 

Summarizing the results, we would like to emphasize that the proposed Richard's curve model describes properly the growth in the number of COVID-19 daily positives and daily deceased, despite its simplicity. Indeed, it is able to reflect properly the trend of the considered daily incidence indicators and also allows for the straightforward inclusion of exogenous information. Basic covariates such as the week-day effect proved to sensibly enhance model fitting and prediction accuracy. 
While we have illustrated results at the national level, the model can clearly be used also at regional/local level (including specific local effects).

The maximum likelihood approach so far considered is rather stable, as long as reasonable starting values are passed to initialize the algorithm. Of course, different approaches could be investigated.
In further work, a Bayesian approach will be experimented in order to overcome possible issues with the asymptotic properties of the maximum likelihood estimator.
Notably, implementation of the \textit{No-U-Turn Sampler} (NUTS) algorithm for the estimation of non-linear models might be a valid working solution. Additionally, a Bayesian approach may also be used to include spatial dependence into the modeling framework and also to relax the first-order Markov assumption for taking into account more complex temporal dependence.
In particular, the latter may be key in order to adapt the introduced Richard's curve model for the nowcasting of prevalence indicators, e.g. \textit{current positives} and \textit{current intensive care units occupancy}. 
Indeed, any modeling effort shall account for the strong temporal dependence between subsequent counts stemming from the fact that daily counts at time $t$ potentially include units which are in stock since times $\tau<t$.
Furthermore, as specified in Section \ref{Subsubsec:Prevalence}, prevalence indicators are non-monotonic and their value is the result of the combination of the incidence components building up each of those. These two last issues may be addressed by adapting the Richard's response function to accommodate non-monotonicity and/or by hierarchically specifying a model for the prevalence indicators through the combination of models for their incidence components. 
A successful attempt in accurately nowcasting the \textit{intensive care units occupancy} is given in \cite{farcomeni2020ensemble}.

\section{Software}
\label{sec6}

Software in the form of R code, together with a sample input data set and complete documentation is available on
request from the corresponding author.

\bibliographystyle{plain}
%\bibliography{template}

\newpage

\section*{Appendix}
\subsection*{Gradients}
In order to make the optimization procedure robust, gradients and Hessians used for the estimation (optimization routine on the log-likelihood) have been computed analytically.
This section provides insights about their derivation for the log-likelihoods at hand. For the sake of clarity, in the sequel, we will invert the previous notation and denote the functions of interest as functions of the parameters, given the observed time points: e.g. $\tilde{\mu}_{\boldsymbol{\theta}}(t)$ becomes $\tilde{\mu}_t(\boldsymbol{\theta})$.
We first provide the computations for the gradient of the log-likelihood for both Poisson and Negative Binomial distributions by considering their mean function $\tilde{\mu}_t(\boldsymbol{\theta})$ as a whole. Afterwards, we show the gradients and introduce the Hessians specific to $\tilde{\lambda}_t(\boldsymbol{\gamma})$, as it is the most cumbersome component of the mean to derive with respect to its parameters.

\subsubsection*{Poisson Gradient}
Let $q$ denote any of the elements of $\boldsymbol{\theta}$, vector of parameters characterizing the mean function $\tilde{\mu}_t(\boldsymbol{\theta})$.
The generic derivative with respect to the component $q$ of $\boldsymbol{\theta}$ for the Poisson log-likelihood $\text{Poi}(\tilde{\mu}_t(\boldsymbol{\theta}))$ is:
\begin{equation}
\begin{aligned}
\frac{\partial}{\partial q}l_{Poi}(\boldsymbol{\gamma}|\mathbf{y})&=-\sum_{t=1}^T\frac{\partial}{\partial q}\tilde{\mu}_t(\boldsymbol{\theta}) + \sum_{t=1}^T y_t\frac{\partial}{\partial q}\log(\tilde{\mu}_t(\boldsymbol{\theta}))=\\
&=-\sum_{t=1}^T\frac{\partial}{\partial q}\tilde{\mu}_t(\boldsymbol{\theta}) + \sum_{t=1}^T y_t\frac{1}{\tilde{\mu}_t(\boldsymbol{\theta})}\frac{\partial}{\partial q}\tilde{\mu}_t(\boldsymbol{\theta}).
\end{aligned}
\label{PoisLlDer}
\end{equation}

\subsubsection*{Negative Binomial Gradient}
The Negative Binomial $\text{NB}\left(\nu, \mu_t(\boldsymbol{\theta})\right)$ presents the additional parameter $\nu$, which does not affect the mean function but controls for the dispersion.
In the following, we provide the first derivative with respect to $\nu$ and with respect to the generic element $q$ of $\boldsymbol{\theta}$, respectively.

The first derivative with respect to $\nu$ of the log-likelihood is:
\begin{equation}
\begin{aligned}
\frac{\partial}{\partial \nu}l_{NB}(\nu, \boldsymbol{\theta}|\mathbf{y})&=T(\log(\nu)-\psi(\nu))+\\
&+\sum_{t=1}^T\left(\psi(\nu+y_t)-log\left(\mu_t(\boldsymbol{\theta})+\nu\right)+\frac{\mu_t(\boldsymbol{\theta})-y_t}{\mu_t(\boldsymbol{\theta})+\nu}\right)
\end{aligned}
\label{NBLlDern}
\end{equation}
where $\psi(\cdot)$ denotes the \textit{digamma} function. \\
The generic derivative with respect to $q$ of the log-likelihood is:
\begin{equation}
\begin{aligned}
\frac{\partial}{\partial q}l_{NB}(\nu, \boldsymbol{\gamma}|\mathbf{y})=&-\sum_{t=1}^T\frac{y_t+\nu}{\tilde{\mu}_t(\boldsymbol{\theta})+\nu}\frac{\partial}{\partial q}\tilde{\mu}_t(\boldsymbol{\theta})+\\
&+\sum_{t=1}^T\frac{y_t}{\tilde{\mu}_t(\boldsymbol{\theta})}\frac{\partial}{\partial q}\tilde{\mu}_t(\boldsymbol{\theta}).
\end{aligned}
\label{NBLlDerGamma}
\end{equation}

\subsection*{Richards' Gradient}
\label{richGrad}
Derivation of the gradient $\tilde{\mu}_t(\boldsymbol{\theta})$ can be obtained by deriving separately (but appropriately) each of the pieces composing it. Computations are straightforward for all components, but for the Richard's first differences parameters, which can in turn be divided as:
\begin{equation*}
    \frac{\partial}{\partial \gamma_i}\tilde{\lambda}_t(\boldsymbol{\gamma})=\frac{\partial}{\partial \gamma_i}\lambda_t(\boldsymbol{\gamma})-\frac{\partial}{\partial \gamma_i}\lambda_{t-1}(\boldsymbol{\gamma})
\end{equation*}

The Richards' function gradient is composed of the following four terms:
\begin{equation*}
\nabla \lambda_t(\boldsymbol{\gamma})=\begin{bmatrix}
% \frac{\partial}{\partial b}\lambda_t(\boldsymbol{\gamma})\\
% \\
\frac{\partial}{\partial r}\lambda_t(\boldsymbol{\gamma})\\
\\
\frac{\partial}{\partial h}\lambda_t(\boldsymbol{\gamma})\\
\\
\frac{\partial}{\partial p}\lambda_t(\boldsymbol{\gamma})\\
\\
\frac{\partial}{\partial s}\lambda_t(\boldsymbol{\gamma})\\
\end{bmatrix}
\end{equation*}
which can be computed as follows.

\begin{equation*}
\begin{aligned}
% 	\frac{\partial}{\partial b}\lambda_t(b) &= \frac{\partial}{\partial b}\left(b+\frac{r}{(1+10^{h(p-t)})^s}\right)=1\\
% 	\\
\frac{\partial}{\partial r}\lambda_t(r) &= \frac{\partial}{\partial r}\left(b+\frac{r}{(1+10^{h(p-t)})^s}\right)=\frac{1}{(1+10^{h(p-t)})^s},\\
\\
\frac{\partial}{\partial h}\lambda_t(h) &= \frac{\partial}{\partial h}\left(b+\frac{r}{(1+10^{h(p-t)})^s}\right)=\\
&=-r\cdot s \cdot (1+10^{h(p-t)})^{-s-1}10^{h(p-t)}(p-t)\log(10),\\
\\
\frac{\partial}{\partial p}\lambda_t(p) &= \frac{\partial}{\partial p}\left(b+\frac{r}{(1+10^{h(p-t)})^s}\right)=\\
&=-r\cdot s \cdot (1+10^{h(p-t)})^{-s-1}10^{h(p-t)}h\log(10),\\
\\
\frac{\partial}{\partial s}\lambda_t(s) &= \frac{\partial}{\partial s}\left(b+\frac{r}{(1+10^{h(p-t)})^s}\right)=\\
&=-r\cdot \left(1+10^{h(p-t)}\right)^{-s} \log\left(1+10^{h(p-t)}\right)
\end{aligned}
\end{equation*}

\subsubsection*{Log-scale}
In the \texttt{R} implementation, the log-likelihood has been parametrized on the log-scale for all the parameters defined on $\mathbb{R}^+$ in order to ease the optimization process under he positivity constraint.
This means that given  $q\in\left\lbrace b, r, p, s\right\rbrace$, the log-likelihood uses  $\log(q)=v$, where $q=e^v$. This implies that, when we do the derivative, we have to take into account the jacobian as a result of the transformation:
\begin{equation}
\begin{aligned}
\frac{\partial}{\partial v}\lambda_t(\boldsymbol{\gamma}) &=\frac{\partial}{\partial e^v}\lambda_t(\boldsymbol{\gamma})\frac{\partial e^v}{\partial v} = \frac{\partial}{\partial e^v}\lambda_t(\boldsymbol{\gamma})\cdot e^v=\frac{\partial}{\partial q}\lambda_t(\boldsymbol{\gamma})\cdot q.
\end{aligned}
\end{equation}
Therefore, each derivative must be multiplied by $e^v=q$.

\section*{Hessians}
Hessians used for the estimation procedure of the model (optimization routine on the log-likelihood) have been computed analytically. In the sequel, we first provide the hessian for the log-likelihod of Poisson and Negative Binomial by considering $\lambda_t(\gamma)$ as a whole.

\subsection*{Poisson Hessian}
Let $q$ and $f$ denote any pair of the parameters characterizing the mean function $\tilde{\mu}_t(\boldsymbol{\theta})$.

The mixed second derivative with respect to the components $q$ and $f$ of $\boldsymbol{\theta}$ for the Poisson log-likelihood is:
\begin{equation*}
\begin{aligned}
\frac{\partial^2}{\partial xf}l_{Poi}(\boldsymbol{\theta}|\mathbf{y})&=\sum_{t=1}^T\frac{y_t-\tilde{\mu}_t(\boldsymbol{\theta})}{\tilde{\mu}_t(\boldsymbol{\theta})}\frac{\partial^2}{\partial q f}\tilde{\mu}_t(\boldsymbol{\theta}) - \sum_{t=1}^T \frac{y_t}{\tilde{\mu}_t(\boldsymbol{\theta})^2}\frac{\partial}{\partial q}\tilde{\mu}_t(\boldsymbol{\theta})\frac{\partial}{\partial f}\tilde{\mu}_t(\boldsymbol{\theta}).
\end{aligned}
\end{equation*}

The second derivative with respect to the components $q$ for the Poisson log-likelihood is:
\begin{equation*}
\begin{aligned}
\frac{\partial^2}{\partial q^2}l_{Poi}(\boldsymbol{\theta}|\mathbf{y})&= \sum_{t=1}^T \frac{y_t-\tilde{\mu}_t(\boldsymbol{\theta})}{\tilde{\mu}_t(\boldsymbol{\theta})}\frac{\partial^2}{\partial q^2} \tilde{\mu}_t(\boldsymbol{\theta}) - \sum_{t=1}^T \frac{y_t}{\tilde{\mu}_t(\boldsymbol{\theta}) ^2}\left(\frac{\partial}{\partial q}\tilde{\mu}_t(\boldsymbol{\theta})\right)^2.
\end{aligned}
\end{equation*}

\subsection*{Negative Binomial Hessian}
Let $q$ and $f$ denote any pair of the parameters characterizing the mean function $\mu_t(\boldsymbol{\theta})$.

The mixed second derivative with respect to $q$ and $f$ of the Negative Binomial log-likelihood is:
\begin{equation*}
\begin{aligned}
\frac{\partial^2}{\partial xf}l_{NB}(\boldsymbol{\theta}|\mathbf{y})&=\sum_{t=1}^T\left(\frac{y_t+\nu}{(\mu_t(\boldsymbol{\theta})+\nu)^2}-\frac{y_t}{\mu_t(\boldsymbol{\theta})^2}\right)\frac{\partial}{\partial q}\tilde{\mu}_t(\boldsymbol{\theta})\frac{\partial}{\partial f}\tilde{\mu}_t(\boldsymbol{\theta})&+\\
&+\sum_{t=1}^T\left(\frac{y_t}{\tilde{\mu}_t(\boldsymbol{\theta})}-\frac{y_t+\nu}{\tilde{\mu}_t(\boldsymbol{\theta})+\nu}\right)\frac{\partial^2}{\partial q f}\tilde{\mu}_t(\boldsymbol{\theta}).
\end{aligned}
\end{equation*}

The second derivative with respect to $q$ of the Negative Binomial log-likelihood is:
\begin{equation*}
\begin{aligned}
\frac{\partial^2}{\partial q^2}l_{NB}(\boldsymbol{\theta}|\mathbf{y})&=\sum_{t=1}^T\left(\frac{y_t+\nu}{(\tilde{\mu}_t(\boldsymbol{\theta})+\nu)^2}-\frac{y_t}{\tilde{\mu}_t(\boldsymbol{\theta})^2}\right)\left(\frac{\partial}{\partial q}\tilde{\mu}_t(\boldsymbol{\theta})\right)^2&+\\
&+\sum_{t=1}^T\left(\frac{y_t}{\tilde{\mu}_t(\boldsymbol{\theta})}-\frac{y_t+\nu}{\tilde{\mu}_t(\boldsymbol{\theta})+\nu}\right)\frac{\partial^2}{\partial q^2}\tilde{\mu}_t(\boldsymbol{\theta}).
\end{aligned}
\end{equation*}

In the Negative Binomial case, we must recall the presence of the additional parameter $\nu$.
The second derivative with respect to $\nu$ of the Negative Binomial log-likelihood is:
\begin{equation*}
\begin{aligned}
\frac{\partial^2}{\partial \nu^2}l_{NB}(\boldsymbol{\theta}|\mathbf{y})&= T\left(\frac{1}{\nu} - \psi^{'}(\nu) \right) + \\ &
+ \sum_{t = 1}^T \left(\psi'(\nu+y_t) - \frac{1}{\tilde{\mu}_t(\boldsymbol{\theta}) + \nu} - \frac{\tilde{\mu}_t(\boldsymbol{\theta}) - y_t}{(\tilde{\mu}_t(\boldsymbol{\theta}) + \nu)^2}\right)
\end{aligned}
\end{equation*}
where $\psi(\cdot)$ and $\psi'(\cdot)$ denote the \textit{digamma} and the \textit{trigamma} function, respectively. \\
The mixed derivative with respect to $\nu$ and the generic element $q$ of $\boldsymbol{\gamma}$ is:
\begin{equation*}
\begin{aligned}
\frac{\partial^2}{\partial \nu q}l_{NB}(\boldsymbol{\gamma}|\mathbf{y})&= \sum_{t = 1}^T\frac{y_t-\tilde{\mu}_t(\boldsymbol{\theta})}{(\tilde{\mu}_t(\boldsymbol{\theta})+\nu)^2}\frac{\partial}{\partial q}\tilde{\mu}_t(\boldsymbol{\theta})
\end{aligned}
\end{equation*}

\subsection*{Richards' Hessian}
As for the gradient, the same holds for the Hessian of the first differences of the Richards function, which would be only one interesting computation to show.
Also here:
\begin{equation*}
    \frac{\partial^2}{\partial \gamma_i\gamma_j}\tilde{\lambda}_t(\boldsymbol{\gamma})=\frac{\partial^2}{\partial \gamma_i\gamma_j}\lambda_t(\boldsymbol{\gamma})-\frac{\partial^2}{\partial \gamma_i\gamma_j}\lambda_{t-1}(\boldsymbol{\gamma})
\end{equation*}
In particular, the resulting Hessian is a $4\times 4$ matrix such that:
\begin{equation*}
\left[\mathbf{H}\left(\lambda_t(\boldsymbol{\gamma})\right)\right]_{ij}=\frac{\partial^2}{\partial  \boldsymbol{\gamma}_i\boldsymbol{\gamma}_j}\lambda_t(\boldsymbol{\gamma}), \quad i,j\in\left\lbrace{1,\dots,4}\right\rbrace.
\end{equation*}
Computations are straightforward for most of the terms, but the final result counts $10$ terms (the Hessian matrix is symmetric) and some of those terms are cumbersome to report. Therefore, we won't include these in the appendix. The reader is invited to contact the authors if he is interested in their detailed computation.

\subsubsection*{Log-scale}
In the \texttt{R} implementation, the log-likelihood has been parametrized on the log-scale for all the parameters defined on $\mathbb{R}^+$ in order to ease the optimization process under the positivity constraint.
This means that given two generic elements, say $q$ and $f$, of the parameters' vector $\boldsymbol{\gamma}$, the log-likelihood uses $\log(q)=v$ and $\log(f)=u$, where $q=e^v$ and $f=e^u$.
The Jacobian inclusion  has two implications on the Hessian.

When computing the mixed derivative, we need to account for the transformation of both terms (if both are on the log scale):
\begin{equation}
\begin{aligned}
\frac{\partial^2}{\partial v\partial u}\lambda_t(\boldsymbol{\gamma}) &= \frac{\partial}{\partial v}\left(\frac{\partial}{\partial u}\lambda_t(\boldsymbol{\gamma})\right)=\frac{\partial}{\partial v}\left(\frac{\partial}{\partial e^u}\lambda_t(\boldsymbol{\gamma})\frac{\partial e^u}{\partial u}\right) =\\
&= \frac{\partial}{\partial v}\left(\frac{\partial}{\partial e^u}\lambda_t(\boldsymbol{\gamma})\cdot e^u\right)=\frac{\partial}{\partial e^v}\left(\frac{\partial}{\partial e^u}\lambda_t(\boldsymbol{\gamma})\cdot e^u\right)\frac{\partial e^v}{\partial v}=\\
&=\frac{\partial}{\partial e^v}\left(\frac{\partial}{\partial e^u}\lambda_t(\boldsymbol{\gamma})\cdot e^u\right)\cdot e^v=\frac{\partial}{\partial q}\frac{\partial}{\partial s}\lambda_t(\boldsymbol{\gamma})\cdot q\cdot f.
\end{aligned}
\end{equation}
Therefore, each mixed derivative $\frac{\partial^2}{\partial xs}\lambda_t(\boldsymbol{\gamma})$ must be multiplied by both $e^v=q$ and $e^u=f$.

When computing the second derivative for $v$, we need to recall that the first derivative contains the jacobian, so:
\begin{equation}
\begin{aligned}
\frac{\partial^2}{\partial v}\lambda_t(\boldsymbol{\gamma}) &= \frac{\partial}{\partial v}\left(\frac{\partial}{\partial v}\lambda_t(\boldsymbol{\gamma})\right)=\frac{\partial}{\partial e^v}\left(\frac{\partial}{\partial e^v}\lambda_t(\boldsymbol{\gamma})\cdot e^v\right)\cdot e^v=\\
&=\left(\frac{\partial}{\partial e^v}\frac{\partial}{\partial e^v}\lambda_t(\boldsymbol{\gamma})\cdot e^v+\frac{\partial}{\partial e^v}\lambda_t(\boldsymbol{\gamma})\cdot \frac{\partial}{\partial e^v}e^v\right)\cdot e^v=\\
&= \left(\frac{\partial}{\partial q}\frac{\partial}{\partial q}\lambda_t(\boldsymbol{\gamma})\cdot q+\frac{\partial}{\partial q}\lambda_t(\boldsymbol{\gamma})\cdot \frac{\partial}{\partial q}q\right)\cdot q=\\
&=\frac{\partial^2}{\partial q^2}\lambda_t(\boldsymbol{\gamma})\cdot q^2+\frac{\partial}{\partial q}\lambda_t(\boldsymbol{\gamma})\cdot q.
\end{aligned}
\end{equation}

\end{document}